\documentclass{jpp}
\usepackage{graphicx}
\usepackage{tensor}
\usepackage{bm}
\usepackage{enumitem}
\usepackage{stackrel}

\usepackage{tikz}
\newcommand*\circled[1]{\tikz[baseline=(char.base)]{
            \node[shape=circle,draw,inner sep=2pt] (char) {#1};}}

\newlist{myitemize}{itemize}{3}
\setlist[myitemize,1]{label=\textbullet,leftmargin=2em,rightmargin=2em,itemindent=0pt,labelsep=5pt,labelwidth=2em}
\setlist[myitemize,2]{label=$\rightarrow$,leftmargin=1em}
\setlist[myitemize,3]{label=$\diamond$}

\newlist{myenumerate}{enumerate}{1}
\setlist[myenumerate,1]{leftmargin=3em,rightmargin=3em,itemindent=0pt,labelsep=5pt,labelwidth=2em}

\usepackage[utf8]{inputenc}
\usepackage[T1]{fontenc}
\usepackage{amsmath}

\renewcommand{\tt}{\tilde{\theta}}							
\renewcommand{\d}{\textrm{d}}								

\makeatletter
\newcommand{\vast}{\bBigg@{3}}
\newcommand{\Vast}{\bBigg@{4}}
\makeatother

\title{The gyrokinetic dispersion relation of microtearing
  modes in collisionless toroidal plasmas}

\author{B.\ D.\ G.\ Chandran\aff{1}\corresp{\email{benjamin.chandran@unh.edu}}
  and A.\ A.\ Schekochihin\aff{2,3}}

\affiliation{\aff{1}Space Science Center and Department of Physics
    and Astronomy,  University of New Hampshire, Durham, NH 03824
\aff{2} Rudolf Peierls Centre for Theoretical Physics, University of
Oxford, Oxford OX1 3NP, United Kingdom
    \aff{3} Merton College, University of Oxford, Oxford OX1 4JD, United Kingdom
}

\begin{document}

\maketitle

\begin{abstract}
  We solve the linearized gyrokinetic equation, quasineutrality
  condition, and Ampere's law to obtain the dispersion relation of
  microtearing modes (MTMs) in collisionless low-$\beta$ toroidal
  plasmas. Consistent with past studies, we find that MTMs are driven
  unstable by the electron temperature gradient and that this
  instability drive is mediated by magnetic drifts. The
  dispersion relation that we derive can be evaluated numerically very
  quickly and may prove useful for devising strategies to mitigate MTM instability in
  fusion devices.
\end{abstract}

\vspace{0.2cm} 
\section{Introduction}
\label{sec:intro}
\vspace{0.2cm}

The free energy stored in magnetically confined, toroidal plasmas
gives rise to a number of  plasma instabilities. As fluctuations
excited by these
instabilities grow, they develop into turbulence, which rapidly
transports thermal energy from the plasma core to the plasma edge. The
resulting heat loss is a major obstacle to developing a commercially
viable fusion reactor, and finding ways to reduce turbulent transport
is one of the primary goals of current fusion research. An important
step towards achieving this goal is to determine the linear stability
thresholds of the relevant plasma modes.

At sufficiently small values of~$\beta$ (the ratio of plasma pressure
to magnetic pressure), it is difficult for plasma fluctuations to
perturb the magnetic field, and the dominant instabilities, such as
the ion- and electron-temperature-gradient modes, are
electrostatic \citep[see, e.g.][]{cowley91,dorland00}. However,
as~$\beta$ increases, electromagnetic instabilities, such as the
microtearing mode~(MTM) and kinetic ballooning mode~(KBM), eventually
become the main drivers of turbulence. Such electromagnetic
instabilities are of particular relevance to spherical tokamaks, in
which~$\beta$ is typically several times larger than in conventional
tokamaks \citep[see, e.g.,][ and references therein]{giacomin23,kennedy23}. The purpose of this paper is to derive the gyrokinetic
MTM dispersion
relation in the collisionless
limit, which is relevant to the hot
plasmas in the cores of existing and planned fusion devices.

We have organized the remainder of this paper as follows. In
Sections~\ref{sec:magneticdriftwaves} through~\ref{sec:determination},
we highlight selected results from the literature and preview the main
steps in our derivation of the MTM dispersion relation, which follows
in detail in Section~\ref{sec:collisionless}. In
Section~\ref{sec:numerical}, we present several numerical examples,
and in Section~\ref{sec:conclusion} we discuss our principal findings
and conclude.

\subsection{Magnetic drift waves}
\label{sec:magneticdriftwaves} 

A simple but useful reference point for understanding the MTM is the
isobaric magnetic drift wave in a plasma in which the equilibrium
magnetic field~$\bm{B}$ is uniform and static and neither the
equilibrium electron pressure~$p_{\rm e}$ nor the fluctuating
quantities vary along~$\bm{B}$. If we neglect electron
inertia, then we can write the component of the electron momentum
equation along the total magnetic field $\bm{B} + \delta \bm{B}$ as
\begin{equation} 
\frac{e n_0}{c}  \frac{\partial }{\partial t}\, \delta A_\parallel   =  
\frac{\delta \bm{B}_\perp}{B}\cdot \nabla p_{\rm e},
\end{equation}
or, equivalently, 
\begin{equation} 
\left(  \frac{\partial }{\partial t} 
  + \bm{v}_{\ast \rm e} \cdot   \nabla \right)\delta A_\parallel = 0 ,
\label{eq:mdw1}
\end{equation} 
where~$e$ is the proton charge, $c$ is the speed of light, 
$\delta A_\parallel = \bm{b} \cdot \delta \bm{A}$,
$\bm{b} = \bm{B}/B$, $\delta \bm{A}$ is the perturbation to the vector
potential,
$\delta \bm{B}_\perp = (\nabla \delta A_\parallel) \times \bm{b}$ is
the component of~$\delta \bm{B}$ perpendicular to~$\bm{B}$, and
$\bm{v}_{\ast \rm e} = - c\bm{B} \times \nabla p_{\rm e} / (en_0 B^2)$ is
the electron diamagnetic drift velocity.  This equation is equivalent
to equation~(D20) of \cite{adkins22} in the limit that
$\lambda \gg d_{\rm e}$, where $\lambda$ is the perpendicular
wavelength and $d_{\rm e}$ is the electron skin
depth. Equation~(\ref{eq:mdw1}) describes magnetic drift waves, in
which $\delta A_\parallel$ is advected at velocity~$\bm{v}_{\ast \rm
  e}$.

\subsection{Ballooning transformation, quasimodes, and mode rational surfaces}
\label{sec:ballooning0} 

Throughout the rest of this paper, we consider axisymmetric toroidal
equilibria and focus on individual Fourier
modes~$\propto \exp\left( i n \zeta - i \omega t\right)$ with
infinitesimal amplitudes, where
$\zeta$ is the toroidal angle, $n$ is the toroidal mode number,
and~$\omega$ is the frequency. In order to enforce rapid spatial variation perpendicular to~$\bm{B}$, slow
variation along~$\bm{B}$, and periodicity in the poloidal angle~$\theta$, we set~$n\gg 1$
and employ the ballooning
transformation \citep{connor78, connor79,tang80}:
\begin{equation}
  \bm{u} (\psi, \theta,\zeta) =
  \sum_{j=-\infty}^\infty \hat{\bm{u}}(\psi,\theta+2\pi j)
  \exp\left\{ in \left[\alpha(\psi, \theta + 2\pi j, \zeta) + \int^\psi \overline{ k}(\psi^\prime)
      {\rm d}\psi^\prime \right]\right\},
    \label{eq:ballooning_intro}
\end{equation}
where~$\bm{u}$ is a vector whose components are the various
fluctuating quantities, $\psi$ is the poloidal flux, and
$\overline{ k}(\psi)$ is a function that is discussed in the run-up
to~(\ref{eq:theta0}). The triad $(\alpha, \psi, \theta)$ is a Clebsch
coordinate system, in which $\alpha(\psi, \theta, \zeta)$ (defined
in~(\ref{eq:defalpha})) and $\psi$ are constant along magnetic-field
lines, while the poloidal angle $\theta$ serves to measure position
along~$\bm{B}$.  Although the (position-space) mode
$\bm{u}(\psi,\theta,\zeta)$ is periodic in~$\theta$, the
(ballooning-space) `quasimode'~$\hat{\bm{u}}$ is not.  Instead,
$\hat{\bm{u}}(\psi, \theta) \rightarrow 0$ as
$\theta \rightarrow \pm \infty$ to ensure that the sum in
(\ref{eq:ballooning_intro}) converges. As discussed further in
Section~\ref{sec:tearing}, the very broad $\theta$ envelope of the
MTM's electrostatic potential eigenfunction~$\delta \hat{\Phi}$
implies that $\delta \Phi$ in position space is peaked around mode
rational surfaces on which~$n q(\psi)$ is an integer, where $q(\psi)$
is the safety factor defined
in~(\ref{eq:defq})~\citep{cowley91,hardman23}. It follows from
(\ref{eq:BpB}) and~(\ref{eq:n_ordering}) that mode rational surfaces
are, for each~$n$, spaced a distance~$\sim k_\wedge^{-1}$ apart, where
$k_\wedge$ is the binormal wavenumber (the wavevector component
perpendicular to both~$\bm{B}$ and~$\nabla \psi$) defined
in~(\ref{eq:kwedge}). Because of magnetic shear (illustrated in the
right panel of figure~\ref{fig:perturbed_field_line}), quasimode
structure at~$|\theta|\gg 1$ corresponds to mode structure at spatial
scales $\sim (k_\wedge |\theta|)^{-1}$ in the~$\nabla \psi$ direction
(see~(\ref{eq:kpsi})) and (\ref{eq:kwedge})).

\subsection{Tearing parity}
\label{sec:tearing0}

\begin{figure}
  \begin{center}
    \hspace{-0.8cm} 
  \includegraphics[width=6cm]{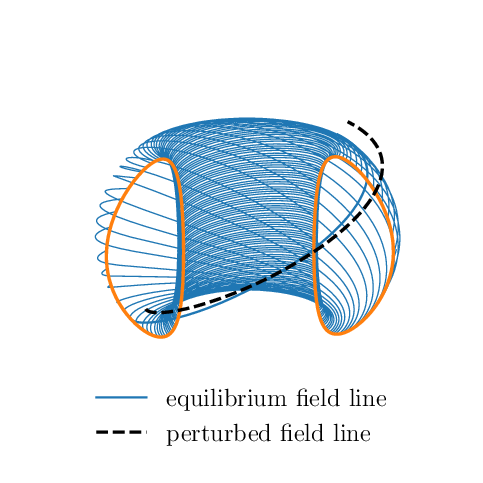}
  \hspace{0.5cm} 
  \includegraphics[width=6cm]{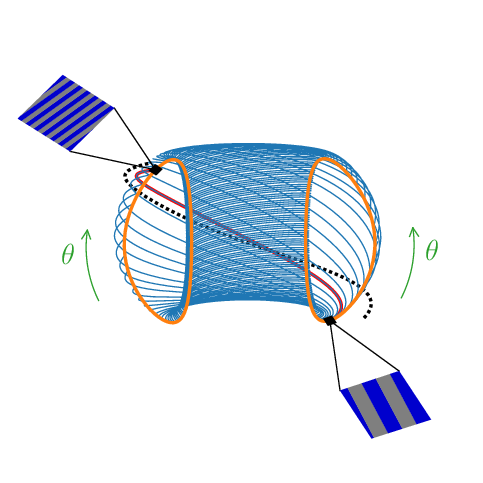}
  \caption{
The blue lines
    are segments of an equilibrium magnetic-field line that
    traces out a mode rational surface in a hypothetical spherical tokamak. 
    {\em Left:} the black dashed curve shows, in an exaggerated
    fashion, how a segment of this
    field line might be perturbed by an MTM.
{\em Right:} the red line
    highlights one of the blue field-line segments, and the black dotted line is a nearby
    equilibrium magnetic-field line at slightly larger~$\psi$. We assume ${\rm
      d}q/{\rm d}\psi >0$, where~$q$ is defined in (\ref{eq:defq}), so
    that the black dotted field line rotates
    through a smaller $\theta$ interval than the solid red line as the
    two lines traverse the same interval of toroidal angle. This
    magnetic shear rotates the phase fronts of the MTMs, causing them
    to draw closer together in the~$\nabla \psi$ direction as one
    follows the red field-line segment from the lower right-hand side of the figure to
    the upper left-hand side, as illustrated schematically by the blue-and-gray-striped squares.
    \label{fig:perturbed_field_line} }
\end{center}
\end{figure}

MTMs involve $\delta A_\parallel$ perturbations that behave like the
magnetic drift waves described
in~Section~\ref{sec:magneticdriftwaves}, propagating at a
velocity~$\simeq \bm{v}_{\ast \rm e}$ (see
figure~\ref{fig:omega_k_wedge}).
As we discuss in greater detail in~Section~\ref{sec:tearing},
a defining feature of the MTM is
`tearing parity,' which means that $\delta \hat{A}_\parallel$ has a
non-vanishing line integral along the magnetic field
\citep{hatch10,dickinson11,ishizawa15,patel22}.  This in turn implies
that, as one follows a perturbed magnetic-field line at a mode rational
surface, the field line wanders secularly in the~$\psi$ direction
\citep{hardman23}, as illustrated in the left half of
figure~\ref{fig:perturbed_field_line}.
Magnetic-field lines perturbed by MTMs thus create channels for electrons
to transport heat  down the
temperature gradient, enabling MTMs to tap into the free energy stored
in the electron temperature profile \citep{drake80,guttenfelder12b}.
In contrast, in KBMs, a
perturbed magnetic-field line at a mode rational surface returns to
its initial equilibrium magnetic flux surface after each poloidal
revolution about the plasma (see Section~\ref{sec:tearing}). This essential difference between
the MTM and KBM is why the MTM (in contrast to the KBM) is driven by
the electron temperature gradient (and the rapid transport of heat
along perturbed magnetic-field lines by electrons) and not by the
density gradient \citep[see Section~\ref{sec:zero_eta_e} and, e.g.,][]{hazeltine75,drake77,hassam80,
  applegate07,predebon13,guttenfelder12a,
  zocco15,hamed19,geng20,patel22,hardman23,yagyu23,giacomin23}.

\subsection{Characteristic scales and quasimode eigenfunctions}
\label{sec:quasimode}

The characteristic MTM binormal wavenumber satisfies
$k_\wedge \rho_{\rm e} \ll 1$, where~$\rho_{\rm e}$ is the electron
gyroradius. As we discuss further in
Section~\ref{sec:collisionless}, the $\delta \hat{A}_\parallel$
fluctuation of an MTM is approximately localized to a $\theta$
interval of order unity \citep[e.g.,][]{applegate07,hamed19,hardman23}.
Because the MTM has tearing parity, this localized
$\delta \hat{A}_\parallel$ fluctuation creates parallel current
density~$\delta \hat{j}_\parallel$ via two very powerful mechanisms:
the rapid streaming of passing electrons along perturbed
magnetic-field lines that connect different equilibrium magnetic flux
surfaces, and the parallel inductive electric field
$-c^{-1} (\partial /\partial t)\delta \hat{A}_\parallel$. In
Section~\ref{sec:conclusion}, we label the current densities created by these two
mechanisms~$\delta \hat{j}_{\delta B_\psi}$
and~$\delta \hat{j}_{\delta E_\parallel}$, respectively, and give
mathematical expressions for each.  Because of the rapid motion of electrons, the non-Boltzmann part of the
perturbation to the passing-electron gyrokinetic distribution
function, denoted by~$\hat{h}_{\rm e, passing}$, created by these two
current-generation mechanisms persists out to great distances along the
magnetic field (i.e., out to $|\theta | \gg 1$) and generates, via the
quasineutrality condition, a perturbation to the electrostatic
potential~$\delta \hat{\Phi}$ that likewise extends to
$|\theta| \gg 1$ \citep{hardman22,hardman23}.

The width of the $\delta \hat{\Phi}$ eigenfunction in~$\theta$ is
ultimately limited by several factors. Magnetic shear
(i.e., non-zero ${\rm d}q/{\rm d}\psi$) 
endows $\delta \hat{\Phi}$ and $\hat{h}_{\rm e, passing}$ at
large~$|\theta|$ with spatial structure at scales~$\ll k_\wedge^{-1}$
in the~$\nabla \psi$ direction, as illustrated in the right panel of
figure~\ref{fig:perturbed_field_line}. In addition, at
large~$|\theta|$, passing electrons at the same position but different
velocities that are moving towards larger~$|\theta|$ will have
previously interacted with
electromagnetic fluctuations at smaller~$|\theta|$ that were at
substantially different phases, which causes
$\hat{h}_{\rm e, passing}$ at sufficiently large~$|\theta|$ to become
a rapidly varying function of velocity.  As we discuss further in
Appendix~\ref{ap:justification}, the rapid variation of
$\hat{h}_{\rm e, passing}$ (in space and velocity) causes
$\delta \hat{\Phi}$ to decay (via gyroaveraging and phase mixing) at
$|\theta| \gtrsim (k_\wedge \rho_{\rm e})^{-1}$
\citep[cf.,][]{hardman22, hardman23}.\footnote{A complex
  frequency also contributes to the truncation of the fluctuations at
  large~$|\theta|$~\citep[see, e.g.,][]{frieman80}.  For example, when
  the growth rate is positive, passing electrons with
  $\pm v_\parallel \bm{\hat{b}} \cdot \nabla \theta > 0$ at
  $ \theta \rightarrow \pm \infty$ will have interacted with electromagnetic
  fluctuations at finite~$\theta$ at very early times when the
  fluctuations were vanishingly small.}
As the MTM $\delta \hat{\Phi}$ eigenfunction extends out to~$|\theta|
\sim (k_\wedge \rho_{\rm e})^{-1}$ in ballooning space, the 
$\delta \Phi$ fluctuations in position space have a characteristic
scale~$ (k_\wedge \theta)^{-1} \sim
\rho_{\rm e}$ in the~$\nabla \psi$ direction.

\subsection{The MTM dispersion relation}
\label{sec:determination}

Our derivation of the MTM dispersion relation in
Section~\ref{sec:collisionless} consists of four steps to determine
the four unknowns $\hat{h}_{\rm e}$,
$\delta \hat{A}_\parallel$, $\delta \hat{\Phi}$, and~$\omega$,
where $\hat{h}_{\rm e}$ is the non-Boltzmann part of the perturbed
electron distribution function for both passing and trapped electrons.
First,
we integrate the gyrokinetic equation~(\ref{eq:gk}) to solve
for~$\hat{h}_{\rm e}$ in terms of~$\delta \hat{A}_\parallel$,
$\delta \hat{\Phi}$, and~$\omega$. Second, we take
$\partial/\partial \theta$ of~$1/B$ times the parallel component of
Ampere's law to show that $\delta \hat{A}_\parallel$ is localized at
$\theta \sim {\rm O}(1)$. This localization implies that further
appearances of $\delta \hat{A}_\parallel$ are always inside its line
integral along the magnetic field,
$\int_{-\infty}^\infty J B \,\delta \hat{A}_\parallel \,{\rm d}
\theta$, where $J = (\bm{B} \cdot \nabla \theta)^{-1}$ is the Jacobian of the $(\alpha, \psi, \theta)$
coordinate system. As long as it is non-zero (as it is for a
tearing-parity mode), this integral can be factored out of the remaining
equations as an overall normalization constant. Third,  we
evaluate the parallel component of Ampere's law at~$\theta=0$ to
obtain a single equation for the two remaining unknowns,~$\omega$
and~$\delta \hat{\Phi}$. Finally, we use the
quasineutrality condition to solve for~$\delta \hat{\Phi}$ in terms
of~$\omega$ and plug this value back into the parallel component of
Ampere's law at~$\theta=0$.

Our analysis is greatly simplified by the two-scale nature of the
problem.  In particular, the contribution of~$\delta \hat{\Phi}$ to
the parallel current at~$\theta=0$, denoted
by~$\delta \hat{j}_{\delta \Phi}$, is dominated by the
$\delta \hat{\Phi}$ fluctuations at
$\theta \sim \left(k_\wedge \rho_{\rm e}\right)^{-1}$. As a
consequence, when we use the
quasineutrality condition to solve for~$\delta \hat{\Phi}$ in step~4
of the programme described in previous paragraph, we can, to leading order, restrict our
attention to values of~$|\theta|$ that are sufficiently large that: (1) the
non-Boltzmann part of the perturbed ion distribution function
$\hat{h}_{\rm i}$ can be neglected because of gyroaveraging and phase mixing,  and (2)
$\delta \hat{A}_\parallel$ enters the quasineutrality condition only via the quantity
$\int_{-\infty}^\infty J B \,\delta \hat{A}_\parallel \,{\rm d}
\theta$, as already mentioned.

We note that $\delta \hat{j}_{\delta \Phi}$ does not arise from the
parallel electric field associated with~$\delta \Phi$, whose effects
are included in the Boltzmann response, which is an 
even function of the parallel velocity and hence does not generate
parallel current.  Instead, $\delta \hat{j}_{\delta \Phi}$ arises from
electron energization or de-energization caused by the partial time
derivative of the electrostatic potential energy
$-e\partial \delta \hat{\Phi}/\partial t$ and from $\delta \hat{\Phi}$
causing electrons to $\bm{E} \times \bm{B}$-drift across the
equilibrium flux surfaces.\footnote{These latter two effects are
  represented mathematically by the terms proportional to~$\omega \delta \hat{\Phi}$
and~$\Omega_{\rm \ast e}(E) \delta
\hat{\Phi}$, respectively, on the right-hand side of~(\ref{eq:gk}).}

\section{Derivation of the MTM dispersion relation}
\label{sec:collisionless}

We consider a gyrokinetic model of a single-ion-species
plasma whose equilibrium state is axisymmetric. A comprehensive
derivation of the equations describing this system is reviewed by
\cite{abel13}, who included plasma rotation, which we neglect for
simplicity. In this model, the equilibrium distribution function of
species~$s$ (with $s = \mathrm{i}$ for ions and $s = \mathrm{e}$ for
electrons) is a Maxwellian, and the number density $n_0$ and
temperature $T_{s}$ are flux functions:
\begin{equation}
  F_{ 0s} = \frac{n_0(\psi)}{\pi^{3/2} v_{\it T \it s}^3}
  \exp\left( - \frac{m_{s} E}{T_{ s}(\psi)} \right),
  \label{eq:F0s}
\end{equation}
where $\psi$ is the poloidal flux, $E = v^2/2$,
$\bm{v}$ is the particle velocity,
$  v_{T  s} = \left( 2 T_{s}/m_{s}\right)^{1/2}$
is the thermal speed of species~$s$, 
and $m_{s}$ is the mass of a particle of species~$s$.

The equilibrium magnetic field $\bm{B}$ can be written
in two equivalent ways:
the standard form for axisymmetric equilibria,
\begin{equation}
  \boldsymbol{B} = \nabla \zeta \times \nabla \psi + I(\psi) \nabla \zeta,
\label{eq:BpsiI} 
\end{equation} 
and the Clebsch form~\citep{kruskal58}
\begin{equation}
\boldsymbol{B} = \nabla \alpha \times \nabla \psi.
\label{eq:Clebsch}
\end{equation} 
Here, $\zeta$ is the toroidal angle,
 $I(\psi)$ is the axial
current divided by~$2\pi$, 
\begin{equation}
  \alpha(\psi, \theta, \zeta) \equiv \zeta - q(\psi) \theta - \nu(\psi, \theta),
  \label{eq:defalpha}
\end{equation}
$\theta$ is the poloidal angle,
\begin{equation} 
q(\psi ) \equiv \frac{1}{2\pi}
\int_0^{2\pi} \frac{\boldsymbol{B} \cdot \nabla \zeta}{\boldsymbol{B} \cdot\nabla \theta} \,{\rm d}\theta 
\label{eq:defq} 
\end{equation} 
is the safety factor, 
and 
\begin{equation}
  \nu(\psi, \theta) = \int_0^\theta \frac{\boldsymbol{B} \cdot \nabla
    \zeta}{\boldsymbol{B} \cdot \nabla \theta^\prime} {\rm d}
  \theta^\prime - q(\psi)\theta.
\label{eq:defnu}
\end{equation}
The $\theta$ integrals in (\ref{eq:defq}) and~(\ref{eq:defnu}) are
evaluated at constant~$\psi$.  Unlike~$\alpha$, $\nu$ is a
single-valued, periodic function of~$\theta$. As mentioned in
Section~\ref{sec:ballooning0}, in Clebsch coordinates $(\alpha, \psi,
\theta)$, $\alpha$ and~$\psi$ serve to label the magnetic-field
lines, and~$\theta$ determines the position along a magnetic-field line.

\subsection{Ballooning transformation}
\label{sec:ballooning} 

As mentioned in Section~\ref{sec:ballooning0}, we represent all
fluctuating quantities as Fourier series in the toroidal angle~$\zeta$
and focus on a single Fourier mode with toroidal mode number~$n$.
Because the spatial variation of MTMs in the plane
perpendicular to~$\bm{B}$ is much more rapid than their spatial variation
along~$\bm{B}$, it would
be natural to take all fluctuating quantities to be of the form
$ f(\psi, \theta) \exp\{i n [\alpha + g(\psi)]\}$ with $|n| \gg 1$,
where $f(\psi, \theta)$ and~$g(\psi)$ are slowly varying functions.
However, as pointed out by \cite{connor78}, fluctuations of this form
are unphysical when~$q$ is irrational, because they are not periodic
in~$\theta$. This is problematic because $q$ is irrational in
essentially all of the plasma volume when $q^\prime(\psi) \neq 0$.

To circumvent this difficulty,
we follow \cite{connor78}, \cite{tang80}, and others by employing
the ballooning transformation,
\begin{equation}
  \bm{u}(\psi, \theta, \zeta)=\sum_{j=-\infty}^\infty
  \hat{\bm{u}}\left(\psi, \theta + 2\pi j\right) e^{i n S(\psi, \theta
    + 2 \pi j, \zeta)},
  \label{eq:ballooning}
\end{equation}
where $ \bm{u}$ is a vector whose components are the various
fluctuating quantities,
$ |n| \gg 1$,   $\hat{\bm{u}}$ is a slowly
varying function of~$\psi$ and~$\theta$, and
$\bm{B} \cdot \nabla S = 0$.
These last three conditions 
guarantee rapid spatial variation, but only in directions
perpendicular to~$\bm{B}$.
We require that
$ \hat{\bm{u}}(\psi, \theta)\rightarrow 0$ sufficiently rapidly as
$|\theta | \rightarrow \infty$ for the sum in~(\ref{eq:ballooning}) to
converge. As mentioned in
Section~\ref{sec:ballooning0}, we refer to~$\bm{u}$ as the `mode'
and~$\hat{\bm{u}}$ as the `quasimode.' The ballooning
transformation represents $\bm{u}$ as the sum
of an infinite number of copies of $\hat{\bm{u}} e^{inS}$ that are
translated in~$\theta$ by successive integer multiples of~$2\pi$,
thereby ensuring that~$\bm{u}$ is periodic in~$\theta$.

As we are taking  $\bm{u}(\psi, \theta, \zeta)$ to be $\propto e^{i n \zeta}$
with no other $\zeta$ dependence, the condition $\bm{B} \cdot \nabla S
= 0$ implies that \citep{tang80}
\begin{equation}
S = \alpha + \int^\psi \overline{k}(\psi^\prime)\,{\rm d}\psi^\prime,
\label{eq:defS} 
\end{equation} 
where $\overline{ k}(\psi)$ is some function of~$\psi$ alone. Thus,
the eikonal form conjectured in the first paragraph of this section
describes the rapid cross-field spatial variation of the summand
in~(\ref{eq:ballooning}) rather than the spatial variation
of~$\bm{u}(\psi, \theta, \zeta)$ in its entirety. The function
$\overline{k}(\psi)$ can in principle be determined through a global
analysis, but here we carry out a local analysis about some flux
surface~$\psi=\psi_0$, with $\overline{ k}(\psi_0)$ a free parameter
that is related to the ballooning angle \citep[see,
e.g.,][]{hardman22}
\begin{equation}
  \theta_0 = \frac{\overline{ k}(\psi_0)}{q^\prime(\psi_0)}.
  \label{eq:theta0}
\end{equation}

If we represent the linear eigenvalue problem that determines the MTM
eigenfunctions and dispersion relation in the form
\begin{equation}
  {\cal L}\,  \bm{u} = 0,
\label{eq:eigenvalue0} 
\end{equation} 
 where ${\cal L}$ is a linear
operator whose coefficients are periodic in~$\theta$ with
period~$2\pi$, then 
the condition
\begin{equation}
  {\cal L} \,   \overline{ \bm{u}} = 0
  \label{eq:quasimode}
\end{equation}
is sufficient for~$  \bm{u}$ 
to solve~(\ref{eq:eigenvalue0}). In subsequent sections, we will
solve~(\ref{eq:quasimode}) rather than~(\ref{eq:eigenvalue0}) 
and simplify notation by writing
$ \hat{\bm{u}}(\psi, \theta) = \hat{\bm{u}}( \theta)$ without
explicitly referencing the slow dependence on~$\psi$.

The perpendicular wavevector is 
\begin{equation}
  \bm{k}_\perp = n \nabla S.
  \label{eq:kperp}
\end{equation}
The component of~$\bm{k}_\perp$ in the~$\nabla \psi$ direction
can be written in the form \citep{hardman22}
\begin{equation}
  k_{\perp \psi} \equiv \bm{k}_\perp \cdot \frac{\nabla \psi}{|\nabla
    \psi|} = n q^\prime(\psi) |\nabla \psi| \left(\theta_0 - \theta\right) -
  \frac{n}{|\nabla \psi|} \left( q \nabla \psi \cdot \nabla \theta +
    \nabla \psi \cdot \nabla \nu\right),
  \label{eq:kpsi}
\end{equation}
which  shows that $|k_{\perp \psi}|$ grows approximately linearly
with~$\theta$ when $|\theta | \gg 1$ in the presence of magnetic shear
(non-zero $q^\prime$). 
The binormal wavenumber is 
\begin{equation}
  k_\wedge = \bm{k}_\perp \cdot \left( \frac{\nabla \psi}{|\nabla
      \psi|} \times \bm{b}  \right)
  = n \nabla \alpha \cdot
  \left( \frac{\nabla \psi}{|\nabla
      \psi|} \times \bm{b}  \right)
  = \frac{n}{|\nabla \psi|} \left(\nabla \alpha \times \nabla \psi\right)
    \cdot \bm{b} = \frac{nB}{|\nabla \psi|},
  \label{eq:kwedge}
\end{equation}
where $\bm{b} = \bm{B}/B$, and the fourth equality in~(\ref{eq:kwedge}) 
follows from~(\ref{eq:Clebsch}).

\subsection{Mode rational surfaces and tearing parity}
\label{sec:tearing} 

With the aid of (\ref{eq:defalpha}) and~(\ref{eq:defS}), we can
rewrite (\ref{eq:ballooning}) in the form
\begin{equation}
  \bm{u}(\psi, \theta, \zeta)=\sum_{j=-\infty}^\infty
  \hat{\bm{u}}(\psi, \theta + 2\pi j) e^{\large i n\left[\alpha(\psi,
      \theta, \zeta) - 2\pi q(\psi)  j + \int^\psi \overline{ k}(\psi^\prime) {\rm d} \psi^\prime\right]}
  \label{eq:ballooning2}\; .
\end{equation}
As discussed in
Section~\ref{sec:intro}, $\delta \hat{\Phi}$ retains a comparable
magnitude as $|\theta|$ increases to values~$\gg 1$. If one were to
treat~$\delta \hat{\Phi}(\theta)$ as approximately constant
out to large values of~$|\theta|$, then the sum on the right-hand side
of~(\ref{eq:ballooning2}) would add to large values  (exhibiting
constructive interference of quasimodes) at mode rational surfaces on
which~$nq(\psi) = m$, where~$m$ is an integer \citep{cowley91}. For
this reason,
in position space, the electrostatic-potential fluctuations of MTMs are peaked
on mode rational surfaces.

Mode rational surfaces have an additional significance related to the perturbed magnetic-field lines. As mentioned
in Section~\ref{sec:tearing0}, MTMs, unlike KBMs, satisfy the 
tearing-parity condition \citep{hatch10,dickinson11,ishizawa15,patel22}
\begin{equation}
\int_{-\infty}^\infty {\rm d}\theta\; J B
\,\delta\hat{A}_{\parallel} \sim
\int_{-\infty}^\infty {\rm d}\theta\; J B
\,|\delta\hat{A}_{\parallel}|,
\label{eq:tearingparity}
\end{equation} 
where
$J = \left[(\nabla \zeta \times \nabla \psi) \cdot \nabla
  \theta\right]^{-1} =
\left[(\nabla \alpha \times \nabla \psi) \cdot \nabla
  \theta\right]^{-1} =
\left(\boldsymbol{B}\cdot \nabla
  \theta\right)^{-1}$ is the Jacobian of both the $(\zeta, \psi, \theta)$
and $(\alpha, \psi, \theta)$ coordinate systems that was previously
mentioned in Section~\ref{sec:determination}.
Equation~(\ref{eq:tearingparity}) implies that
perturbed magnetic-field lines at mode rational surfaces wander secularly towards either
larger or smaller~$\psi$ \citep{hardman23}. To show this, we parameterize the
perturbed magnetic-field line that passes through position
$(\alpha_1, \psi_1, \theta_1)$ using the Clebsch
coordinate functions $\alpha(\theta) = \alpha_1 + \delta \alpha(\theta)$ and
$\psi(\theta) = \psi_1 + \delta \psi(\theta)$, with $\delta \alpha(\theta_1) =
0$ and~$\delta \psi(\theta_1) = 0$. We
define
$l(\theta)$ to be the distance along this perturbed magnetic-field line and $s_\perp(\theta)$ to be the
distance between this perturbed field line and the equilibrium flux
surface $\psi = \psi_1 $.  To leading order in the (infinitesimal) MTM amplitude, 
\begin{equation} 
\begin{split}
\delta \psi(\theta_1 + 2\pi) 
  &= \int_{\theta_1}^{\theta_1+2\pi} \frac{{\rm d}\psi}{{\rm d}s_\perp}
  \frac{{\rm d}s_\perp}{{\rm d}l}
  \frac{{\rm d}l}{{\rm d}\theta}
  \,{\rm d}\theta \\
&  =
  \int_{\theta_1}^{\theta_1+2\pi} |\nabla \psi| 
     \sum_{j=-\infty}^\infty\delta \hat{B}_\psi(\theta+ 2\pi
     j)e^{in(S_1 - 2\pi j q_1 )} \frac{{\rm d}\theta}{\bm{B} \cdot \nabla \theta},
    \label{eq:DeltaPsi2}
\end{split}
\end{equation} 
where we have taken~$s_\perp$ to increase in the direction of
increasing~$\psi$ and~$l$ to increase in the direction
of~$\bm{\hat{b}}$, $\delta \hat{B}_\psi = \delta \hat{\bm{B}} \cdot
\nabla \psi / |\nabla \psi|$,
$S_1 = \alpha_1 + \int^{\psi_1}\overline{ k}(\psi) {\rm d}\psi$, $q_1
= q(\psi_1)$, and
the $\theta$~integral  in~(\ref{eq:DeltaPsi2}) is carried out
at~$\alpha = \alpha_1$ and~$\psi = \psi_1$.
 To leading order in~$1/n$,
$\delta \hat{B}_\psi = - i k_\wedge \delta \hat{A}_\parallel$. If we
take~$\psi=\psi_1$ to be a mode rational surface on which~$nq_1$
is an integer, then, with the aid
of~(\ref{eq:kwedge}), we can rewrite~(\ref{eq:DeltaPsi2}) as
\begin{equation}
\begin{split}
\delta \psi(\theta_1+2\pi)\Big|_{n q_1 =
  \mathrm{integer}} &= - i n e^{inS_1} \sum_{j=-\infty}^\infty
\int_{\theta_1}^{\theta_1+2\pi}
J(\theta) B(\theta) \,\delta \hat{A}_\parallel(\theta + 2\pi j) \, {\rm d}\theta
\\
&= - i n e^{inS_1} \int_{-\infty}^\infty J B \,\delta
\hat{A}_\parallel\,
{\rm d} \theta.
\end{split}
\label{eq:DeltaPsi3} 
\end{equation}
Equation~(\ref{eq:tearingparity}) implies that the right-hand side of
(\ref{eq:DeltaPsi3}) is non-zero. (In contrast,
$\int_{-\infty}^\infty J B \,\delta \hat{A}_\parallel\, {\rm d} \theta$
vanishes for KBMs in the low and intermediate-frequency
regimes \citep{tang80}.) Because the right-hand side of (\ref{eq:DeltaPsi3}) is a
function of~$\alpha_1$ and~$\psi_1$ but not~$\theta_1$, a perturbed
magnetic-field line on a mode rational surface keeps wandering in the
same direction in~$\psi$ each time it winds around the plasma in the
poloidal direction.

\subsection{Orderings}
\label{sec:ordering} 

We assume that
\begin{equation}
 \beta_{\rm
    e}  \equiv \frac{8 \pi n_0 T_{\rm e}}{B^2} \ll 1
  \label{eq:beta_ordering}
\end{equation}
and that
\begin{equation}
  \frac{B_{\rm p}}{B}  \sim \frac{a}{R} \sim q(\psi) \sim a |\nabla q|
  \sim {\rm O}(1),
  \label{eq:BpB} 
\end{equation}
where $B_{\rm p}$ is the
poloidal magnetic field, $a$ is the plasma minor radius, and $R$ is
the plasma major radius.
We take the mode's frequency~$\omega$ to satisfy
\begin{equation}
|\omega| \sim
  |\omega_{\ast \rm e}| ,
  \label{eq:omegaordering}
\end{equation}
where
\begin{equation}
  \omega_{\ast  s} =   n \frac{c T_{ s}}{Z_{ s}e} \frac{\d
    \ln n_0}{\d \psi}
  \label{eq:omegastar}
  \end{equation} 
  is the diamagnetic drift frequency of species~$s$,
  $Z_{ s} e$ is the charge of species~$s$,
$e$ is the proton charge, and $c$ is the speed of light.
We also assume that
\begin{equation}
  k_\wedge \rho_{\rm e} \ll 1,
  \label{eq:kwedge_lim} 
\end{equation} 
where $\rho_{\rm e} = v_{\it T\rm e} / |\Omega_{\rm e}|$
is the electron gyroradius, and
$\Omega_{s} = Z_{\rm s} eB/(m_{s} c)$
is the cyclotron frequency of species~$s$.
We note that, 
from~(\ref{eq:kwedge}), (\ref{eq:BpB}), and~(\ref{eq:omegastar}),
\begin{equation}
  n \sim k_\wedge a \qquad \mbox{and} \qquad \omega_{\ast \rm e} \sim k_\wedge
  \rho_{\rm  e} \frac{v_{\it T\rm e}}{a}.
  \label{eq:n_ordering}
\end{equation}

\subsection{Linearized gyrokinetic equation}
\label{sec:gke} 

In the limit of infinitesimal fluctuation
amplitudes, the ballooning-space representation of
the non-Boltzmann, gyrotropic part of the
perturbed gyrokinetic distribution function
$\hat{h}_{ s}$ 
satisfies the linearized gyrokinetic equation~\citep{tang80},
\begin{equation} 
v_\parallel (\boldsymbol{b}\cdot \nabla\theta) 
\frac{\partial \hat{h}_{ s}}{\partial \theta} 
- i(\omega - \omega_{\rm D \it s}) \,\hat{h}_{ s} 
 = 
-\frac{i Z_{ s} e}{T_{ s}} F_{0 s} \left[ \omega - \Omega_{\ast  s}(E)
\right] J_0(\alpha_{ s}) \left(\delta \hat{\Phi} - \frac{v_\parallel}{c} \delta \hat{A}_\parallel\right),
\label{eq:gk} 
\end{equation}
where $v_\parallel = \bm{v} \cdot \boldsymbol{b}$,
\begin{equation}
  \omega_{\rm D \it s}= \boldsymbol{k}_\perp \cdot \boldsymbol{v}_{\rm D \it s}
\label{eq:omegaDs} 
\end{equation} 
is the magnetic drift frequency,
\begin{equation}
  \boldsymbol{v}_{\rm D \it s} = \frac{\boldsymbol{b}}{\Omega_s}
  \times \left( v_\parallel^2 \boldsymbol{b} \cdot \nabla
    \boldsymbol{b} + \frac{1}{2} v_\perp^2 \nabla \ln B
    \right)
  \label{eq:vD}
\end{equation}
is the guiding-centre drift velocity,
$J_l$ denotes the Bessel function of the first kind of
order~$l$, $\alpha_{s} = k_\perp v_\perp /\Omega_s$
(not to be confused with the Clebsch coordinate~$\alpha$),
\begin{equation}
\Omega_{\ast  s}(E)
= \omega_{\ast  s} \left[ 1 + \eta_{ s}
    \left( \frac{m_{ s}E}{T_{ s}} - \frac{3}{2}
 \right)\right],
  \label{eq:omegastarT}
\end{equation}
and $\eta_s = \d \ln T_s /\d \ln n_0$.
In~(\ref{eq:gk}), the partial derivative~$\partial/\partial \theta$ is taken
at constant $\psi$, $\alpha$, $\mu$, and $E$, where $\mu = v_\perp^2 / (2B)$,
and $v_\perp$ is the velocity component perpendicular to~$\bm{B}$.
In writing~(\ref{eq:gk}), we 
neglected a term involving the parallel component of the fluctuating
magnetic field, which leads to only a small correction to the MTM
dispersion relation
when~$\beta_{\rm e} \ll 1$ \citep{applegate07,patel22,kennedy23}.

\subsection{Passing electrons}
\label{sec:passing} 

To determine $\hat{h}_{\rm e}$ for passing electrons, we
solve~(\ref{eq:gk}) subject to the boundary condition
\begin{equation}
  \lim_{|\theta| \rightarrow \infty}  \hat{h}_{\rm e}(\theta) = 0,
\label{eq:bc}
\end{equation}
which, as noted in section~\ref{sec:ballooning}, is required in order for the sum in~(\ref{eq:ballooning}) to
converge.  The unique solution for $\mathrm{Im }\, \omega > 0$ is given
by \citep{frieman80,tang80}
\begin{equation}
  \hat{h}_{\rm e, passing \pm} = \mp i \xi_{\rm e} \int_{\mp \sigma_J \infty}^\theta {\rm d}\theta^\prime\,
JB J_0(\alpha_{\rm e}) \left(\frac{\delta \hat{\Phi}}{|v_\parallel|}
  \mp \frac{\delta \hat{A}_\parallel }{c}\right)
e^{ \pm i (I_0^\theta - I_0^{\theta ^\prime})}.
\label{eq:integratingfactor}
\end{equation}
Here and in the following, the $\pm$ sign indicates the sign
of~$v_\parallel$,  $\sigma_J = J/|J|$,
$J$ is the Jacobian defined following (\ref{eq:tearingparity}),
\begin{equation} 
  \xi_{ s} \equiv \frac{Z_{ s} e}{T_{ s}} \left[\omega -
\Omega_{\ast  s}(E)\right]F_{0 s},
\label{eq:defxi}
\end{equation} 
and
\begin{equation}
  I_a^b \equiv \int_a^b {\rm d}\theta \, \frac{JB}{|v_\parallel|}(\omega
  - \omega_{\rm De})
  \label{eq:defI}
\end{equation}
is ($-|v_\parallel|/v_\parallel$ times) the change in the MTM
phase factor $nS - \omega t$ at the position of a passing electron as it propagates 
from $\theta=a$ to~$\theta=b$.
In~(\ref{eq:integratingfactor}) and~(\ref{eq:defI}), the
$\theta^\prime$ and $\theta$ integrals are carried out at
constant~$\psi$, $\alpha$, $E$ and~$\mu$.  In (\ref{eq:integratingfactor}) and in the
following, if a function of~$\theta$ appears in an integral
over~$\theta^\prime$ but the function's arguments are not listed, the
function is to be evaluated at~$\theta^\prime$ rather than~$\theta$.
The lower limit of integration in~(\ref{eq:integratingfactor}) is
chosen to ensure that $\hat{h}_{\rm e \pm} \rightarrow 0$
as~$ \theta \rightarrow \mp \sigma_J \infty$. The condition
$\mathrm{Im } \,\omega > 0$ ensures that
$\hat{h}_{\rm e \pm} \rightarrow 0$
as~$\theta \rightarrow \pm  \sigma_J \infty$ because
$\delta \hat{\Phi}(\theta)$ and $\delta \hat{A}_\parallel(\theta)$
also vanish as $|\theta| \rightarrow \infty$. 

\subsection{Leading-order parallel component of Ampere's law and its~$\theta$
  derivative at $\theta \sim \mathrm{O}(1)$}
\label{sec:Ampere} 

In ballooning space, the parallel component of Ampere's law is  \citep{tang80}
\begin{equation}
  \frac{k_\perp^2 c}{4\pi} \delta \hat{A}_\parallel = \delta \hat{j}_\parallel= \sum_{ s} 2\pi Z_s e
\int_0^\infty {\rm d} E \,\int_0^{E/B} {\rm d}\mu\; B \left(
\hat{h}_{ s+} -\hat{h}_{ s-}\right) 
J_0(\alpha_{ s}).
\label{eq:amperegeneral}
\end{equation} 
We only need to evaluate~(\ref{eq:amperegeneral}) 
at~$\theta \sim \mathrm{O}(1)$.
The ion
contribution to the parallel current can be neglected as it
is~$\sim (m_{\rm e}/m_{\rm i})^{1/2}$ times the passing-electron
contribution, the ions being much slower than the
electrons.  The trapped-electron contribution
to the parallel current can also be neglected, as we show in
Appendix~\ref{ap:trapped}.
Using the solution~(\ref{eq:integratingfactor})
in~(\ref{eq:amperegeneral}) and dividing by~$B$, we obtain
\begin{equation} 
\begin{split}
\stackrel{\circled{1}}{\frac{k_\perp^2 c}{4\pi B} \,\delta \hat{A}_\parallel }
\,= \,- &2\pi i e \int_0^\infty {\rm d}E\, \int_0^{E/B_{\rm max}}{\rm d}\mu\,
 J_0(\alpha_{\rm e}) \xi_{\rm e}\vast[  \\
&\int_{-\infty}^\theta {\rm d}\theta^\prime \;|J| B J_0(\alpha_{\rm e}
)\vast(-\sigma_J \frac{\stackrel{\circled{2a}}{\delta \hat{\Phi}}}{|v_\parallel|} \,+\,
  \frac{\stackrel{\circled{3a}}{\delta \hat{A}_\parallel}}{c}\vast)e^{     i
  \left(\overline{I}_0^\theta - \overline{I}_0^{\theta^\prime}\right)} \\
\;\;+\;\; 
&\int_{\theta}^\infty {\rm d}\theta^\prime \;|J| B J_0(\alpha_{\rm e})
\vast(\sigma_J \frac{\stackrel{\circled{2b}}{\delta
    \hat{\Phi}}}{|v_\parallel|} \,+\, \frac{
  \stackrel{\circled{3b}}{\delta \hat{A}_\parallel}}{c}\vast)e^{ -i\left(
\overline{I}_0^\theta - \overline{I}_0^{\theta^\prime}\right)}\,
\vast],
\end{split}
\label{eq:ampere0}
\end{equation} 
where $  \overline{I}_a^b = \sigma_J I_a^b $,
$B_{\rm max}$ is the maximum value of the magnetic field on the
flux surface, and the upper limit of integration of the~$\mu$ integral
restricts the integral to the passing region of velocity space.  The
circled numbers in (\ref{eq:ampere0}) are shorthand for the values of
the terms underneath them after all multiplications and integrations have
been carried out. In Appendix~\ref{ap:justification}, we will show that
\begin{equation}
  \begin{split}
 \circled{2a}  \sim  \circled{2b}  \sim &\left( \circled{3a}  +  \circled{3b}\right)  \\
       \circled{1} \sim \frac{k_\wedge \rho_{\rm e}}{\beta_{\rm e}} \times &\left( \circled{3a} +
          \circled{3b}\right)
\end{split}
\label{eq:comp} 
\end{equation}
at~$\theta \sim \mathrm{O}(1)$, 
a set of relations that we will use in our derivation of~(\ref{eq:divjpar}).

As a first step towards solving (\ref{eq:ampere0}), we consider its
$\theta$ derivative. When
$\partial / \partial \theta$ acts upon the right-hand side
of~(\ref{eq:ampere0}), the resulting quantity is the sum of three
terms.  The first results from taking the $\theta$ derivative of
$J_0(\alpha_{\rm e}(\theta))$, which is
$\propto J_1(\alpha_{\rm e}(\theta))$; this term is a
factor~$\sim k_\wedge \rho_{\rm e}$ smaller than the right-hand
side of~(\ref{eq:ampere0}), because
$\alpha_{\rm e}(\theta) \sim k_\wedge \rho_{\rm e}$ when
$\theta \sim \mathrm{O}(1)$.  The second term results from taking the
$\theta$ derivative of
$\exp\left(\pm i \overline{ I}_0^{\theta}\right)$; it follows
from~(\ref{eq:defI}) that this term is a
factor~$\sim k_\wedge \rho_{\rm e}$ smaller than the right-hand
side of~(\ref{eq:ampere0}) when~$\theta \sim {\rm O}(1)$.  The third term results from evaluating
the integrands in  $\circled{2a}$, $\circled{2b}$, $\circled{3a}$
and $\circled{3b}$ at the endpoints of the $\theta^\prime$
integrations. The resulting $\delta \hat{A}_\parallel$ terms vanish
identically. The resulting $\delta \hat{\Phi}$ terms are a
factor of~$\sim k_\wedge \rho_{\rm e}$ smaller than the sum of
$\circled{2a}$ and~$\circled{2b}$ because, as we will show in
Appendix~\ref{ap:justification}, the $\theta^\prime$
integrand in $\circled{2a}$ has a similar
magnitude throughout the interval
$0 < |\theta| \lesssim (k_\wedge \rho_{\rm e})^{-1}$ before decaying at
larger~$|\theta|$, and likewise for the $\theta^\prime$ integrand in~$\circled{2b}$. 
The integrands in terms~$\circled{2a}$
and $\circled{2b}$ are therefore of order~$\sim k_\wedge \rho_{\rm
  e}$ times the integrals.  To summarize, at
$\theta \sim \mathrm{ O}(1)$, the $\theta$ derivative of the
right-hand side of 
(\ref{eq:ampere0}) is of order~$\sim k_\wedge \rho_{\rm e}$ times
the right-hand side of~(\ref{eq:ampere0}).

In contrast, taking the $\theta$ derivative of the left-hand side of
(\ref{eq:ampere0}) does not change its order
in~$\beta_{\rm e}$ and~$k_\wedge
\rho_{\rm e}$. Hence, when we take the  $\theta$ derivative
of~(\ref{eq:ampere0}) and make use of~(\ref{eq:comp}),
we find that the left-hand side of~(\ref{eq:ampere0}) becomes~$\sim
\beta_{\rm  e}^{-1}$ times larger than the right-hand side, so that,
to leading order,
\begin{equation}
  \frac{\partial}{\partial \theta} \left(\frac{k_\perp^2 c}{4\pi B}
    \delta \hat{A}_{\parallel } \right)  \; = \; 0,
  \label{eq:divjpar}
  \end{equation} 
where the $\theta$ derivative is computed at constant~$\alpha$ and~$\psi$.
Equation~(\ref{eq:divjpar})  is equivalent to the condition that
$\bm{B} \cdot \nabla (\delta \hat{j}_{\parallel } / B)  = \nabla \cdot
(\delta \hat{j}_{\parallel } \bm{b}) = 0$. Upon integration, (\ref{eq:divjpar}) yields
\begin{equation}
     \delta \hat{A}_{\parallel } = C_1 \,\frac{4\pi B}{k_\perp^2 c} ,
\label{eq:c1} 
\end{equation} 
where $C_1$ is a function of~$\psi$, but not of~$\theta$ \citep[cf.][]{hamed19}.

Equations~(\ref{eq:kpsi}) and~(\ref{eq:c1}) imply that
$\delta \hat{A}_{\parallel} \propto \theta^{-2}$ at
large~$|\theta|$; i.e., $\delta \hat{A}_{\parallel}$ is effectively
localized near $\theta \sim \mathrm{O}(1)$, as mentioned
previously. Therefore, the dominant contributions to the
$\theta^\prime$-integrals of~$\delta \hat{A}_\parallel(\theta^\prime)$ appearing
in (\ref{eq:ampere0}) arise from $|\theta^\prime| \sim \mathrm{O}(1)$,
where $\exp\left( \pm i \overline{ I}_0^{\theta^\prime}\right)$ and
$J_0(\alpha_{\rm e}(\theta^\prime))$ are both $\simeq 1$, and, to
leading order in~$\beta_{\rm e}$ and~$k_\wedge \rho_{\rm e}$,
\begin{equation}
  \int_{-\infty}^\theta {\rm d} \theta^\prime \; |J|BJ_0(\alpha_{\rm
    e}) \frac{\delta \hat{A}_{\parallel }}{c} e^{ - i \overline{I}_0^{\theta^\prime}}
+
\int_{\theta}^\infty {\rm d} \theta^\prime \; |J|BJ_0(\alpha_{\rm e})
\frac{\delta \hat{A}_{\parallel }}{c} e^{  i \overline{I}_0^{\theta^\prime}}
= \frac{1}{c} \int_{-\infty}^\infty{\rm d}\theta^\prime \; |J| B
\delta \hat{A}_{\parallel},
\label{eq:psi0pre} 
\end{equation} 
which is independent of $E$ and~$\mu$.  When
$|\theta| \sim \mathrm{O}(1)$,
$\exp\left( \pm i \overline{I}_0^{\theta}\right)$ and
$J_0(\alpha_{\rm e}(\theta))$ equal unity plus small corrections, and,
to leading order in $\beta_{\rm e}$ and~$k_\wedge \rho_{\rm e}$, (\ref{eq:ampere0}) becomes
 \begin{equation}
   \omega - \omega_0 + i \pi^{1/2} \left(\frac{v_{\it T\rm e}}{L}+
     \frac{\omega^2 B_{\rm max}}{2 v_{\it T\rm e}} \int_{-\infty}^\infty \d \theta\,
     |J| \Gamma \, \delta \tilde{\Phi} \right) = 0
   \label{eq:disp0},
\end{equation}
where 
\begin{equation}
  \omega_0 = \omega_{\ast \rm e}\left(1 + \frac{\eta_{\rm e}}{2}\right),
  \qquad
L \; =\; \int_{-\infty}^\infty {\rm d}\theta \; \frac{|J| B^2 \beta_{\rm e}}{B_{\rm max}\left( k_\perp \rho_{\rm e}\right)^2},
\label{eq:beta_e_min}
\end{equation} 
\begin{equation}
\Gamma(\theta) \;= \; \mathrm{sgn}(\theta) \int_{\rm passing} {\rm d}^3 v \;
\frac{F_{0\rm e}}{n_0} \left[\frac{\omega -
\Omega_{\ast \rm e}(E)}{\omega}\right]
J_0(\alpha_{\rm e})\; e^{  i\, \mathrm{sgn}(\theta) \overline{I}_0^\theta},
\label{eq:defGamma} 
\end{equation}
$\int_{\rm passing} {\rm d}^3 v$ is an integral over the
part of velocity space corresponding to passing particles,
\begin{equation}
  \delta \tilde{\Phi} = \frac{\delta
    \hat{\Phi}}{\hat{\psi}_{\parallel,\infty}},
\label{eq:Phi_tilde} 
\end{equation} 
and
\begin{equation} 
\hat{\psi}_{\parallel, \infty} =
\frac{i\omega}{2c}\int_{-\infty}^\infty {\rm d}\theta\; J B
\,\delta\hat{A}_{\parallel}.
\label{eq:defpsi0}
\end{equation} 
In writing (\ref{eq:Phi_tilde}), we
have invoked~(\ref{eq:tearingparity}), which guarantees that
$\hat{\psi}_{\parallel, \infty}$ is non-vanishing.
We note that (\ref{eq:kpsi}) and~(\ref{eq:BpB}) imply that 
$L \sim a \beta_{\rm e}/ (k_\wedge \rho_{\rm e})^2$.
In the next
section, we will show how to determine $\delta \tilde{\Phi}(\theta)$
for any given value of~$\omega$; i.e., how to solve for the function~$\delta
\tilde{\Phi}(\theta, \omega)$. With that solution, (\ref{eq:disp0}) will become
the MTM dispersion relation, which we will solve using Newton's method
in Section~\ref{sec:numerical}.

\subsection{Quasineutrality: determining $\delta \hat{\Phi}_0$  at $|\theta| \gg 1$}
\label{sec:quasineutrality} 

In ballooning space, the quasineutrality condition is~\citep{tang80}
\begin{equation}
  0 = -\sum_{ s}\frac{n_0 Z_{ s}^2 e^2}{T_{ s}}\delta \hat{\Phi} + \sum_{ s} 2\pi Z_{ s} e \int_0^\infty {\rm d}E\,\int_0^{E/B} {\rm d}\mu\;\frac{B}{|v_\parallel|}\left(\hat{h}_{ s+} + \hat{h}_{ s-}\right) J_0(\alpha_{ s}),
  \label{eq:quasineutrality}
\end{equation}
where the first term on the right-hand side
of~(\ref{eq:quasineutrality}) is the Boltzmann response.  Our goal in
this section is to use~(\ref{eq:quasineutrality}) to
determine~$\delta \hat{\Phi}(\theta)$ for any given value
of~$\omega$ so that we can evaluate the $\theta$~integral in
(\ref{eq:disp0}). As shown in Appendix~\ref{ap:justification}, this
integral is dominated by~$|\theta|\sim (k_\wedge \rho_{\rm e})^{-1}$.
The contribution from~$\hat{h}_{\rm i}$ to (\ref{eq:quasineutrality}) 
at such large values of~$|\theta|$ is much smaller than the ion
Boltzmann term because of gyroaveraging: at
$\theta \sim (k_\wedge \rho_{\rm e})^{-1}$, $\alpha_{\rm i} \sim
(m_{\rm i}/m_{\rm e})^{1/2}$ and~$J_0(\alpha_{\rm i})\ll
1$.   With the aid of (\ref{eq:integratingfactor}), we obtain
\begin{equation}
\begin{split}
\left(\hat{h}_{\rm e+} + \hat{h}_{\rm e-} \right)_{\rm passing} =\;\; &
  i \xi_{\rm e} \left[
    \int_{-\infty}^\theta {\rm d} \theta^\prime\, |J| B J_0(\alpha_{\rm
      e}) \left(-\frac{\delta \hat{\Phi}}{|v_\parallel|} + \sigma_{
      J} \frac{\delta \hat{A}_\parallel}{c} \right) e^{i \overline{
        I}_0^\theta - i \overline{ I}_0^{\theta^\prime}}   \right.\\
& \left. -    \int_{\theta}^\infty {\rm d} \theta^\prime \,|J| B J_0(\alpha_{\rm
  e}) \left( \frac{\delta \hat{\Phi}}{|v_\parallel|} + \sigma_{
    J} \frac{\delta \hat{A}_\parallel}{c} \right) e^{-i \overline{
        I}_0^\theta + i \overline{ I}_0^{\theta^\prime}}
    \right].
    \label{eq:h_even}
\end{split}
\end{equation}
Upon substituting (\ref{eq:h_even})   into (\ref{eq:quasineutrality}), 
neglecting~$\hat{h}_{\rm i}$, and
making use of the results in Appendix~\ref{ap:trapped} for the value
of~$\hat{h}_{\rm e}$ for
trapped electrons, we
find that
\begin{equation}
  \delta \tilde{\Phi}(\theta) + \int_{-\infty}^\infty {\rm d}\theta^\prime\,
  W_{\rm  p} \left( \theta,  \theta^\prime\right)  \delta \tilde{\Phi}\left(\theta^\prime\right)
+ \int_{\pi(2j-1)}^{\pi (2j + 1)} {\rm d}\theta^\prime\, W_{\rm tr}\left(\theta, \theta^\prime\right) \delta \tilde{\Phi}\left(\theta^\prime\right)
=  \frac{\tau \Gamma(\theta)}{1 + \tau},
\label{eq:QN}
\end{equation} 
where $\tau = T_{\rm i}/T_{\rm e}$, $j$ is the largest integer such
that~$\pi(2j-1)<\theta$ (we assume that $B$ attains its maximum value at
$\theta=\pi$),
\begin{equation} 
  W_{\rm p}\left( \theta, \theta^\prime\right)\; = \; \frac{2 \pi i \tau}{1+\tau}\left| J(\theta^\prime)\right|
  \int_0^\infty {\rm d}E\, Q(E)  \int_0^{E/B_{\rm max}} {\rm d}\mu \, 
g_+(\theta_{>}, E, \mu) g_-( \theta_{<}, E, \mu),
\label{eq:defW} 
\end{equation}
\begin{equation}
  W_{\rm tr}\left(\theta, \theta^\prime\right) = - \,\frac{4\pi
    \tau}{1 + \tau} \left|J(\theta^\prime)\right| \int_0^\infty {\rm d}E\, Q(E)
  \int_{E/B_{\rm max}}^{\mu_{\rm max}(\theta, \theta^\prime)}
  {\rm d}\mu\, \frac{g_{\rm tr}(\theta, E, \mu) g_{\rm
      tr}(\theta^\prime, E, \mu)}{\left\langle \omega - \omega_{\rm
        De}\right\rangle_{\rm b} \tau_{\rm b}},
\label{eq:Wtr} 
\end{equation}
\begin{equation}
Q(E) = \frac{F_{0\rm e}}{n_0}\left[ \omega - \Omega_{\ast \rm e}(E)\right],
\label{eq:defQ}
\end{equation}
\begin{equation}
  g_\pm ( \theta, E, \mu) = \frac{B J_0(\alpha_{\rm e} ) e^{\pm i \overline{ I}_0^\theta}
   }{|v_\parallel|},
  \qquad
  g_{\rm tr}(\theta, E, \mu) = \frac{B J_0(\alpha_{\rm e})}{|v_\parallel|}  \cos\left(\frac{n q^\prime(\psi) I(\psi)
      |v_\parallel|\theta}{\Omega_{\rm e}} \right),
\label{eq:defg}  
\end{equation}
$\mu_{\rm max}(\theta, \theta^\prime) = {\rm min}[E/B(\theta), E/B(\theta^\prime)]$,
$\theta_> = \max(\theta, \theta^\prime)$,
$\theta_< = \min(\theta, \theta^\prime)$, and $\langle \dots \rangle_{\rm
  b}$ and~$\tau_{\rm b}$ are, respectively, the bounce average and
bounce time defined in~(\ref{eq:tau_b}).
Equation~(\ref{eq:QN})
can be solved numerically by discretizing the integral
over~$\theta^\prime$, which converts~(\ref{eq:QN}) into a matrix
equation for~$\delta \tilde{\Phi}$. We present examples of such solutions in
figure~\ref{fig:Phi}.\footnote{We note that if the equilibrium has
  up-down symmetry and $\theta_0=0$, then $\omega_{\rm De}$ is an even
  function of~$\theta$, $I_0^\theta$ and $\Gamma(\theta)$ are odd
  functions of~$\theta$,
  $W_{\rm p}(-\theta, - \theta^\prime) = W_{\rm p}(\theta,
  \theta^\prime)$,
  $W_{\rm tr}(-\theta, - \theta^\prime) = W_{\rm tr}(\theta,
  \theta^\prime)$, and
  (\ref{eq:QN}) implies that $\delta \hat{\Phi}$ is an odd function
  of~$\theta$, as illustrated in figure~\ref{fig:Phi}.}

We note that~(\ref{eq:disp0}) and~(\ref{eq:QN}) can be combined into a
single eigenvalue equation for~$\delta \hat{\Phi}$ and~$\omega$ by multiplying both equations
by~$\hat{\psi}_{\parallel, \infty}$, using (\ref{eq:disp0}) to solve
for~$\hat{\psi}_{\parallel, \infty}$, and substituting that value
into~(\ref{eq:QN}) to obtain
\begin{equation}
  \delta \hat{\Phi}(\theta) + \int_{-\infty}^\infty {\rm d}\theta^\prime\,
  \overline{ W}_{\rm p} \left( \theta,  \theta^\prime\right) \delta \hat{\Phi}\left(\theta^\prime\right)
+ \int_{\pi(2j-1)}^{\pi (2j + 1)} {\rm d}\theta^\prime\, W_{\rm tr}\left(\theta, \theta^\prime\right) \delta \hat{\Phi}\left(\theta^\prime\right)
= 0,
\label{eq:QN2}
\end{equation} 
where
\begin{equation}
\overline{ W}_{\rm p}(\theta, \theta^\prime) = W_{\rm p}(\theta,
\theta^\prime) + \frac{ i \pi^{1/2} B_{\rm max} \tau \omega^2
  \left|J(\theta^\prime)\right| \Gamma(\theta)
  \Gamma(\theta^\prime)}{2 v_{\it T\rm e} (1 + \tau)\left(\omega -
    \omega_0 + i \pi^{1/2} v_{\it T\rm e}/L\right)}.
\label{eq:olW}
\end{equation} 
Although (\ref{eq:QN2}) could be used to
determine~$\omega$, in this work we obtain the MTM dispersion relation
from~(\ref{eq:disp0}) and~(\ref{eq:QN}).

\subsection{Maximum unstable binormal wavenumber}
\label{sec:krange}

Given the scaling estimates in Appendix~\ref{ap:justification},
the first term~($\omega$), second term~($\omega_0$), and fourth
term~(containing~$\delta \tilde{\Phi}$) in~(\ref{eq:disp0}) are all
comparable, and
the ratio of the third term~($i \pi^{1/2} v_{\it T\rm e}/L$) to these
other terms is~$k_\wedge \rho_{\rm e} / \beta_{\rm e}$. These
estimates also follow from~(\ref{eq:comp}) upon noting that the
$i \pi^{1/2} v_{\it T\rm e}/L$ term in~(\ref{eq:disp0}) comes from
term~$\circled{1}$ in (\ref{eq:ampere0}), the 
$\omega$ and~$\omega_0$ terms come from the sum of
terms~$\circled{3a}$ and~$\circled{3b}$ in~(\ref{eq:ampere0}), and
the  term containing~$\delta \tilde{\Phi}$ in~(\ref{eq:disp0})  comes
from the sum of terms~$\circled{2a}$ and~$\circled{2b}$ in~(\ref{eq:ampere0}).
As the 
$i \pi^{1/2} v_{\it T\rm e}/L$ term in~(\ref{eq:disp0})  is always
stabilizing, the MTM can only be unstable if \citep[cf.][]{hardman23}
\begin{equation}
  k_\wedge \rho_{\rm e} \lesssim \beta_{\rm e}.
  \label{eq:unstable_k}
\end{equation}

\subsection{The limit of long wavelength and cold ions}
\label{sec:cold}

In this section, we specialize our results to the limit in which
\begin{equation}
  \frac{T_{\rm i}}{T_{\rm e}} \sim \frac{k_\wedge \rho_{\rm e}}{\beta_{\rm e} } \ll 1.
  \label{eq:cold_ordering}
\end{equation}
As $W_{\rm p}(\theta, \theta^\prime) $ and~$W_{\rm tr}(\theta, \theta^\prime)$ are $\propto \tau $ when~$\tau \equiv
T_{\rm i}/T_{\rm e}\ll 1$, the
approximate magnitudes of $W_{\rm p}(\theta, \theta^\prime)$
and~$W_{\rm tr}(\theta, \theta^\prime)$ in the
cold-ion limit are the same 
as estimated in Appendix~\ref{ap:justification}, only multiplied
by~$\tau $.
To leading order in~$\tau$, (\ref{eq:QN})  thus yields
\begin{equation}
  \delta \tilde{\Phi} = \tau \Gamma.
  \label{eq:Phi_cold}
\end{equation}
Upon substituting (\ref{eq:Phi_cold}) into (\ref{eq:disp0}) and
defining~$\omega_{\rm r}$ and~$\gamma$ to be the real and imaginary
parts of~$\omega$, respectively,  we find that,  to leading order in~$\tau$,
\begin{equation}
  \omega_{\rm r} = \omega_0
  \label{eq:omega_r_cold}
\end{equation}
and
\begin{equation}
  \gamma = -  \pi^{1/2}  \left[ \frac{v_{\it T\rm e}}{L} + \frac{\tau
      \omega_0^2 B_{\rm max}}{2 v_{\it T\rm e}} \int_{-\infty}^\infty \d \theta\,
     |J|\, {\rm Re}\left(\Gamma^2\right)_{\omega \rightarrow \omega_0} \right],
\label{eq:gamma_cold} 
\end{equation} 
where the subscript `$\omega \rightarrow \omega_0$' means
that~$\omega$ is replaced with~$\omega_0$ in (\ref{eq:defI})
and~(\ref{eq:defGamma}) when
evaluating~$\Gamma$.

\subsection{The limit of long wavelength and zero temperature gradient}
\label{sec:zero_eta_e} 

In this section, we assume that
\begin{equation}
  k_\wedge \rho_{\rm e} \ll \beta_{\rm e}, \qquad \eta_{\rm e} = 0.
  \label{eq:limit_eta_e}
\end{equation}
When $k_\wedge \rho_{\rm e} \ll \beta_{\rm e}$, we can
drop the $i\pi^{1/2} v_{\it T\rm e}/L$ term in~(\ref{eq:disp0}), and
when
$\eta_{\rm e}= 0$, $\Omega_{\ast \rm e}(E) = \omega_{\ast
  \rm e}$. Equation~(\ref{eq:disp0}) then has the solution $\omega =
\omega_{\ast \rm e}$, and for this solution  $\omega - \Omega_{\rm \ast e}(E)$,
$\Gamma(\theta)$, $W_{\rm p}(\theta, \theta^\prime)$, $W_{\rm tr}(\theta, \theta^\prime)$, and~$\delta
\Phi(\theta)$ all vanish. The stability of the MTM at $k_\wedge
\rho_{\rm e} \ll \beta_{\rm e}$
when~$\eta_{\rm e}=0$ implies that the MTM at $k_\wedge \rho_{\rm e}
\ll \beta_{\rm e}$ is driven by the electron
temperature gradient and not by the density gradient.

\section{Numerical examples}
\label{sec:numerical}

\begin{figure}
\begin{center}
  \includegraphics[width=2.5cm]{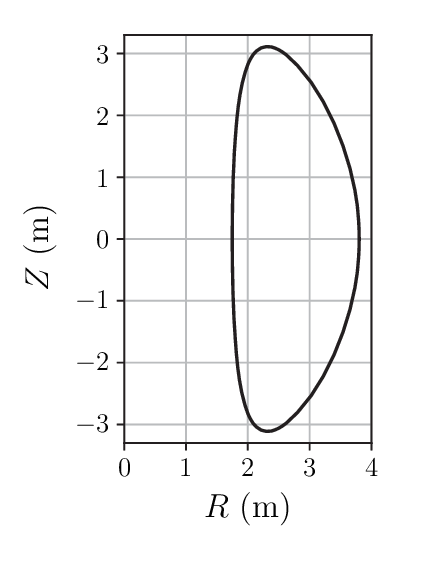}
\hspace{0.1cm} 
  \includegraphics[width=5cm]{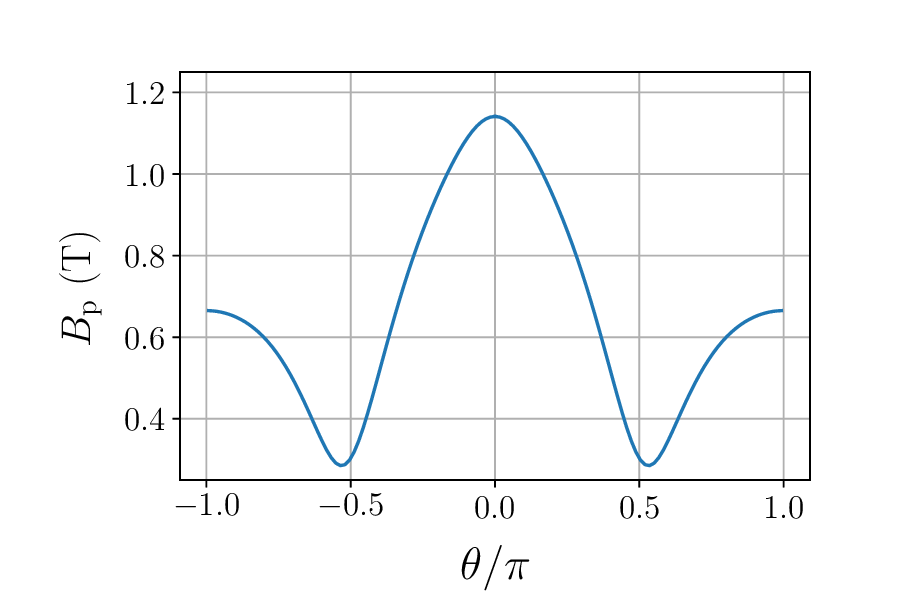}
  \includegraphics[width=5cm]{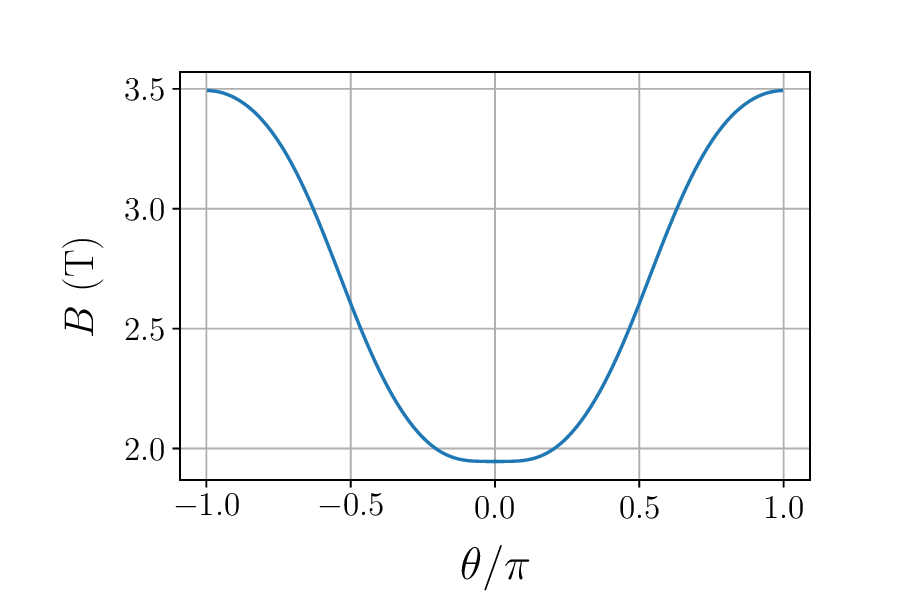}
  \caption{The left, middle, and right plots show, respectively,
    the shape of the flux surface about
  which the local Grad-Shafranov equilibrium is calculated in
  Section~\ref{sec:numerical},
 the strength of the poloidal magnetic field~$B_{\rm p}$ on this flux
 surface as a function of the poloidal angle~$\theta$,
 and the strength of the total magnetic field~$B$ on this flux surface
 as a function of~$\theta$.
    \label{fig:flux_surface} }
\end{center}
\end{figure}

We consider a Miller-Mercier-Luc model \citep{mercier74,miller98} of a
local Grad-Shafranov equilibrium with parameters taken from Table~2 of
\cite{patel22}, except for the electron collision
frequency~$\nu_{\rm e}$, which we set equal to zero. These parameters
were chosen to model a flux surface in the core of the proposed STEP
spherical tokamak~\citep{wilson20}. We plot the shape of this flux
surface and the~$\theta$ profiles of the poloidal and total
magnetic-field strengths on this surface in
figure~\ref{fig:flux_surface}. Throughout this section, we
take~$\theta$ to be the particular poloidal angle used
by~\cite{miller98}. The linear average of
$\beta_{\rm e}$ around the flux-surface contour in the poloidal plane
in this equilibrium is
\begin{equation}
  \beta_{\rm e, av} = 0.125.
\label{eq:beta_av} 
\end{equation} 

To obtain the MTM dispersion relation for this equilibrium, we
solve~(\ref{eq:disp0}) using Newton's method. At each step in
Newton's method, we need to determine~$\delta \tilde{\Phi}$
in~(\ref{eq:disp0}) for some
assumed value of~$\omega$. To do so, we discretize
in~$\theta$, which converts~(\ref{eq:QN}) into a matrix equation
for~$\delta \tilde{\Phi}$. When we solve this matrix equation, we first compute
$\overline{ I}_0^\theta = \sigma_J I_0^\theta$ from (\ref{eq:defI}) on
a $\theta$ grid with 512 uniformly spaced grid points per $2\pi$ increment
in~$\theta$.
We evaluate all integrals over $\theta$,
$E$, and~$\mu$ using the trapezoid rule.
We evaluate $\overline{ I}_0^\theta$ and all other
functions of~$E$ and~$\mu$ on a grid of 64 uniformly spaced grid
points along the velocity~($v$) axis ranging from $0.1 v_{\it T\rm e}$
to $6 v_{\it T\rm e}$ and, at each~$v$, 64 evenly spaced grid points
in~$\mu$ within the passing region of velocity space and another 64
evenly spaced grid points in~$\mu$ within the trapped region of
velocity space.  We evaluate $\Gamma( \theta)$,
$W_{\rm p}(\theta, \theta^\prime)$, $W_{\rm tr}(\theta,
\theta^\prime)$, and~$\delta \tilde{\Phi}$
on a coarser~$\theta$ grid with only 32 points per $2\pi$ increment
in~$\theta$. We adjust the total width of the~$\theta$ grid as we
vary~$n$ to ensure that~$\delta\hat{\Phi}$ decays to small values
before the edge of the grid is reached. In all the examples 
in this section, we set $  \theta_0 = 0$.

\begin{figure}
  \begin{center}
  \includegraphics[width=9cm]{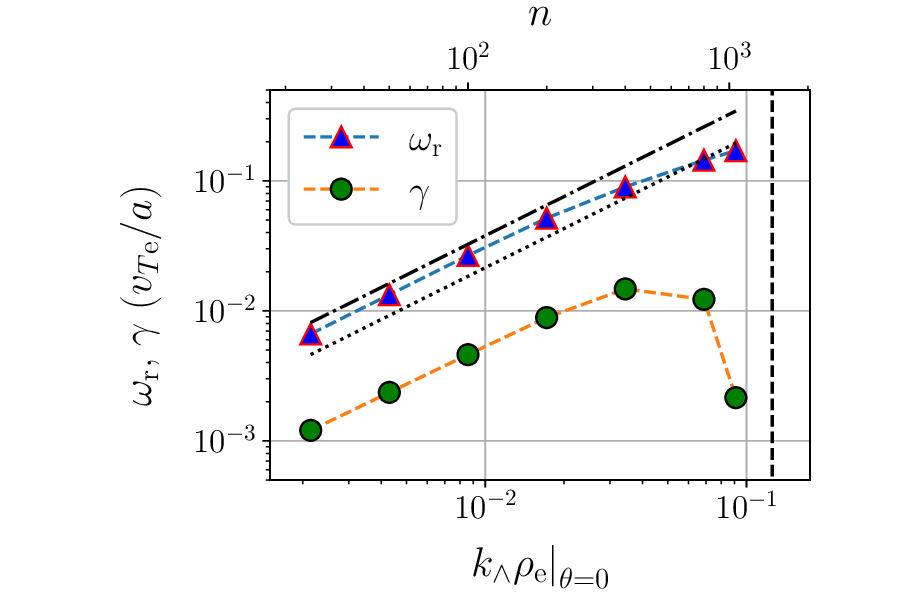}
  \caption{The real and imaginary parts of the MTM frequency
    ($\omega_{\rm r}$ and~$\gamma$, respectively) as a
    function of the toroidal mode number~$n$ (top axis) and
    $k_\wedge \rho_{\rm e}\big|_{\theta=0}$ (bottom axis), where~$k_\wedge$ is the
    binormal wavenumber defined in (\ref{eq:kwedge}).
 Frequencies are given in units of~$v_{\it T\rm e}/a$,
    where~$a$ is the plasma minor radius.
The dotted line is a plot of~$\omega_0$, which is defined in~(\ref{eq:beta_e_min}), and the dash-dotted line is a plot of the
magnetic-drift-wave frequency $\omega_{\rm mdw} = \omega_{\ast \rm
  e}(1+ \eta_{\rm e})$ that follows from (\ref{eq:mdw1}). The vertical
dashed line shows the approximate instability threshold $k_\wedge \rho_{\rm
  e} \lesssim \beta_{\rm e}$ from~(\ref{eq:unstable_k}), where we have set
$\beta_{\rm e}$ equal to the value in~(\ref{eq:beta_av}).
As in all the numerical examples in this paper, we have set the ballooning angle~$\theta_0$ 
equal to zero.
\label{fig:omega_k_wedge} }
\end{center}
\end{figure}

In figure~\ref{fig:omega_k_wedge}, we plot the real and imaginary parts
of~$\omega$, denoted by $\omega_{\rm r} $ and $\gamma$, respectively,
for seven values of~$n$: 25, 50, 100, 200, 400, 800, and 1060. At
$n< 800$, $\omega_{\rm r}$ lies between the cold-ion MTM
frequency~$\omega_0 = \omega_{\ast \rm e}(1 + \eta_{\rm e}/2)$ and the
magnetic-drift-wave frequency that follows from~(\ref{eq:mdw1})
and~(\ref{eq:kwedge}), which is
$\omega_{\rm mdw} = \bm{k}_\perp \cdot \bm{v}_{\ast \rm e} =
\omega_{\ast \rm e}(1+ \eta_{\rm e})$, where the electron diamagnetic
drift velocity~$\bm{v}_{\rm \ast e}$ is defined
below~(\ref{eq:mdw1}). The MTM frequency differs
from~$\omega_{\rm mdw}$ because (\ref{eq:mdw1}) is the 
fluid-theory requirement for avoiding infinite current in a perfectly conducting
plasma with massless electrons, whereas
(\ref{eq:disp0}) is the gyrokinetic statement of the parallel
component of Ampere's law, which accounts for the finite values of the
current and electron mass, the current produced by~$\delta \hat{\Phi}$, and how an
electron's response to the fluctuating fields depends upon the
electron's velocity.  Nevertheless, the approximate
equality~$\omega \simeq \omega_{\rm mdw}$ indicates that the MTM phase
velocity is approximately~$\bm{v}_{\rm \ast e}$ and that the MTM
approximates the force balance that arises in a fluid-theory magnetic
drift wave. Across these same $n$ values ($n< 800$), the MTM growth
rate is approximately 1/6 to 1/5 of~$\omega_{\rm r}$. However, as $n$
increases above~800, $k_\wedge$ approaches the approximate maximum
unstable binormal wavenumber $\beta_{\rm e}/\rho_{\rm e}$ given
in~(\ref{eq:unstable_k}), and $\gamma$ drops sharply.

\begin{figure}
  \begin{center}
  \includegraphics[width=9cm]{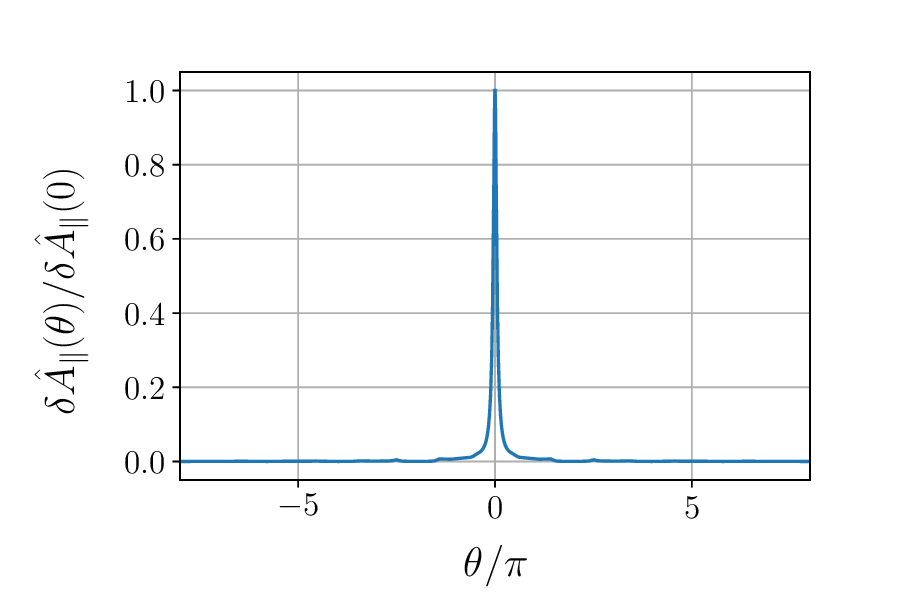}
  \caption{The parallel component
    of the fluctuating vector potential
    $\delta \hat{A}_\parallel(\theta)$ from~(\ref{eq:c1}) 
    normalized to its value at~$\theta=0$ when the ballooning
    angle~$\theta_0$ is zero. The $\theta$ profile of
    $\delta \hat{A}_\parallel(\theta)$ that follows from
    (\ref{eq:c1}) is independent of~$n$.
    \label{fig:Apar} }
\end{center}
\end{figure}

In figure~\ref{fig:Apar}, we plot $\delta \hat{A}_\parallel$
from~(\ref{eq:c1}). We note that the $\theta$ profile
of~$\delta \hat{A}_\parallel$ in~(\ref{eq:c1}) is independent of~$n$,
and thus this same profile applies to all of the data points in figure~\ref{fig:omega_k_wedge}.

\begin{figure}
  \begin{center}
      \includegraphics[width=5.5cm]{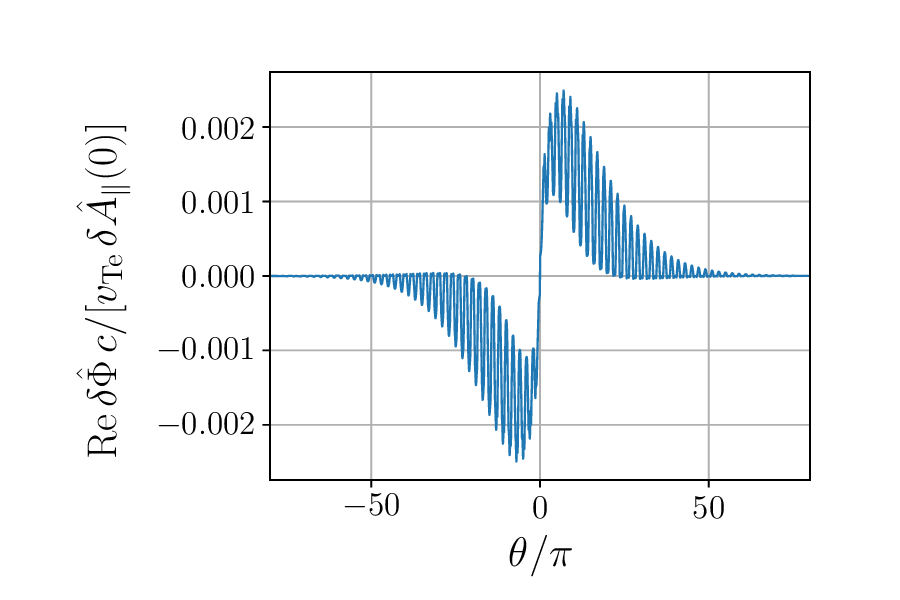}
      \includegraphics[width=5.5cm]{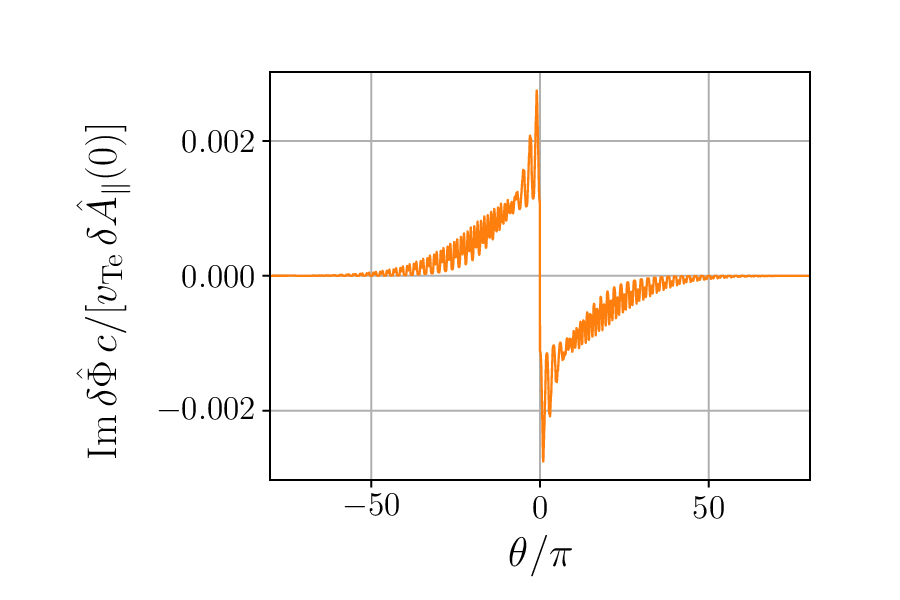}\\
      \includegraphics[width=5.5cm]{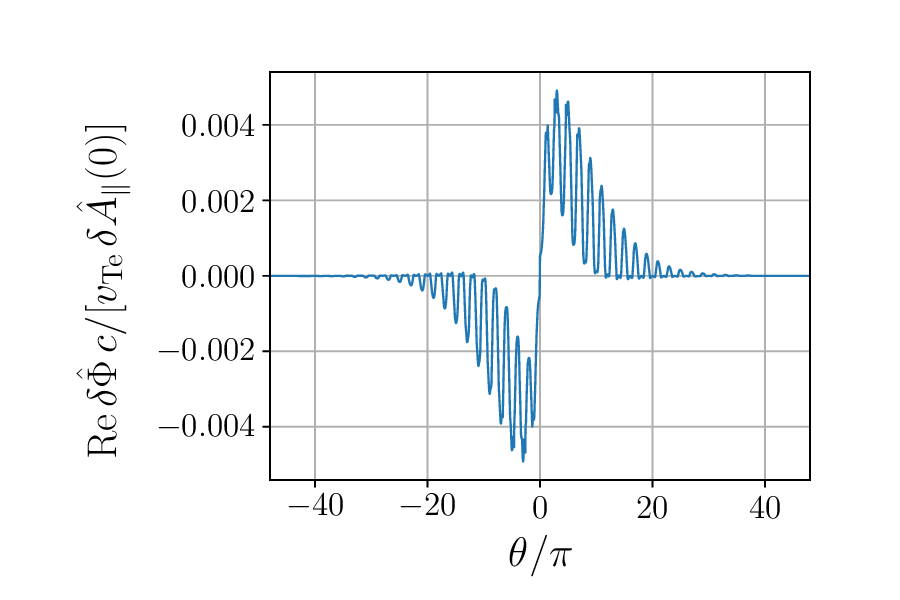}
      \includegraphics[width=5.5cm]{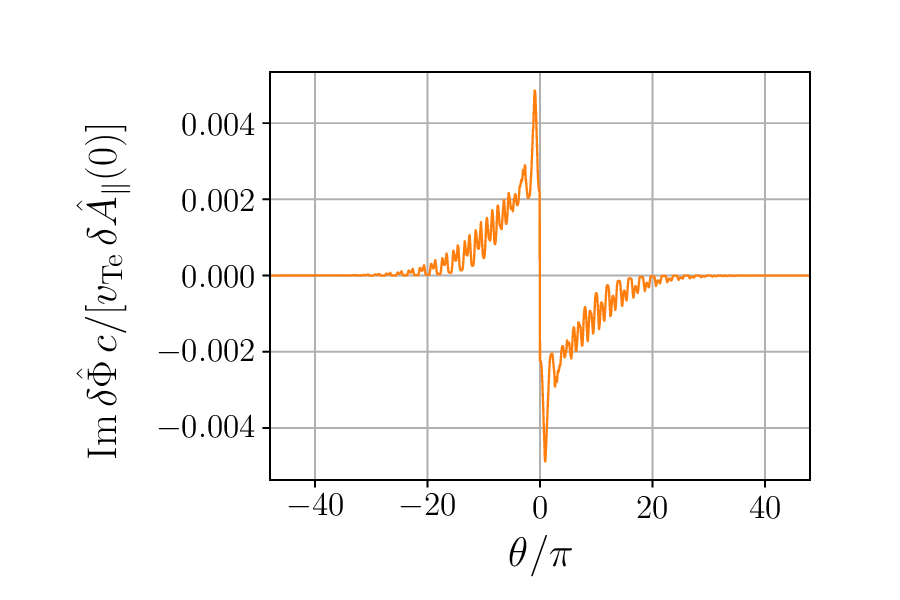}\\
      \includegraphics[width=5.5cm]{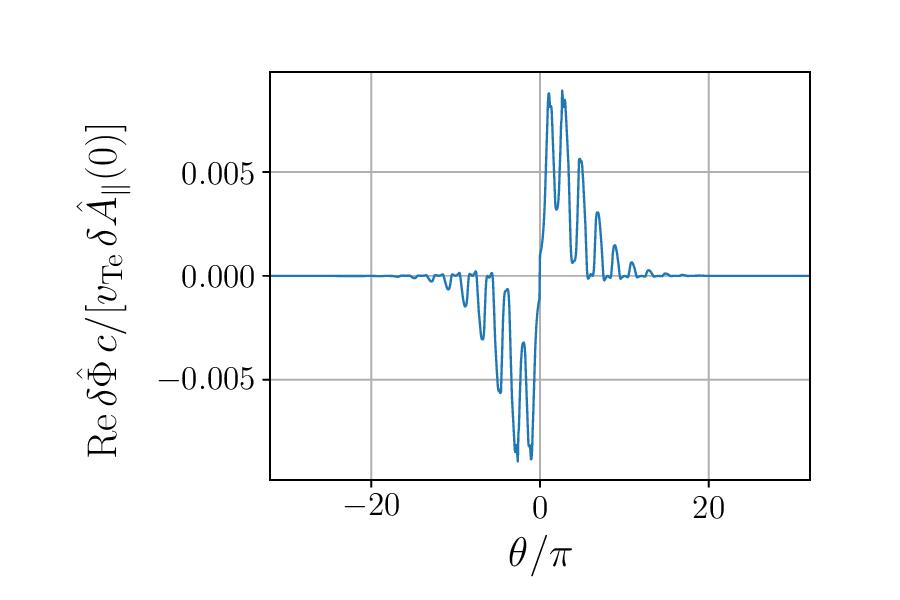}
      \includegraphics[width=5.5cm]{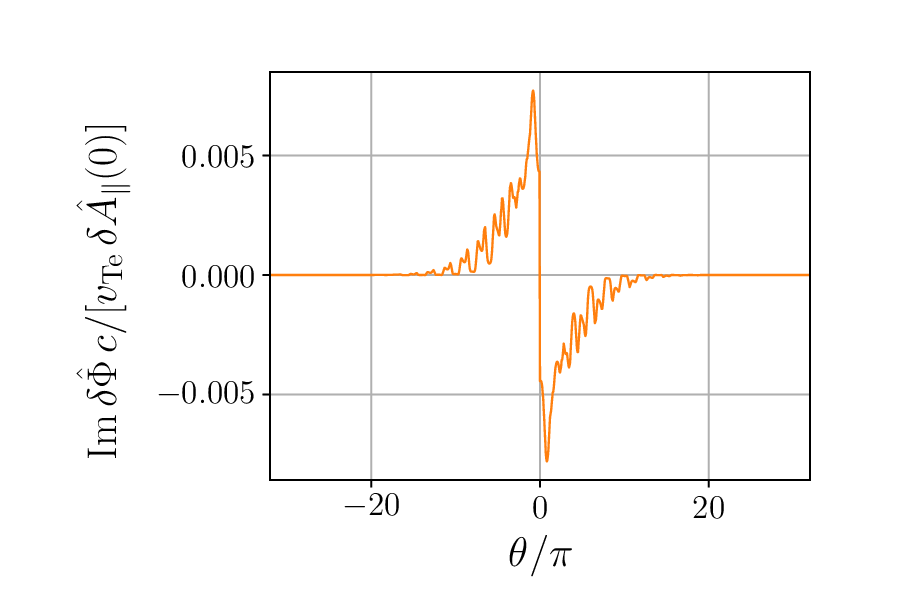}     \\
      \includegraphics[width=5.5cm]{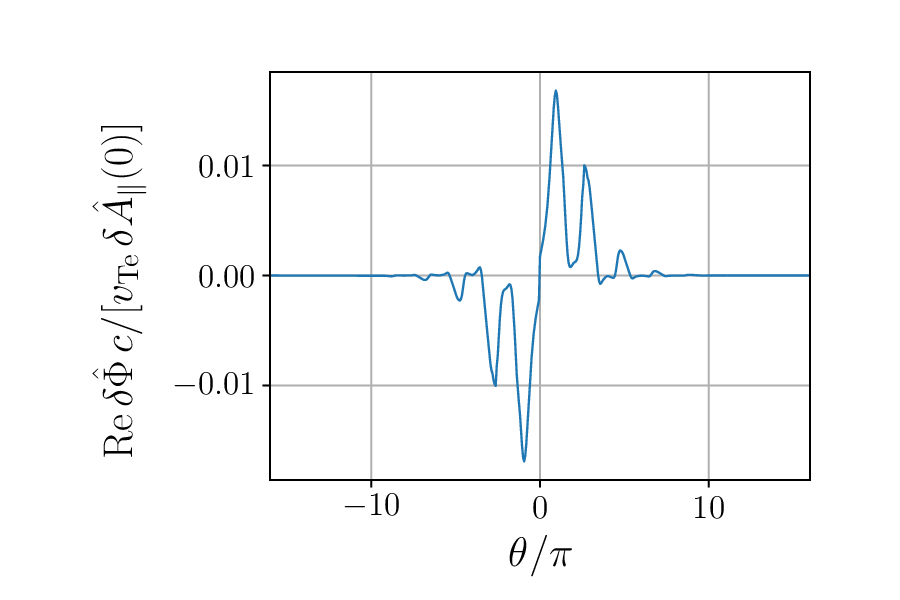}
      \includegraphics[width=5.5cm]{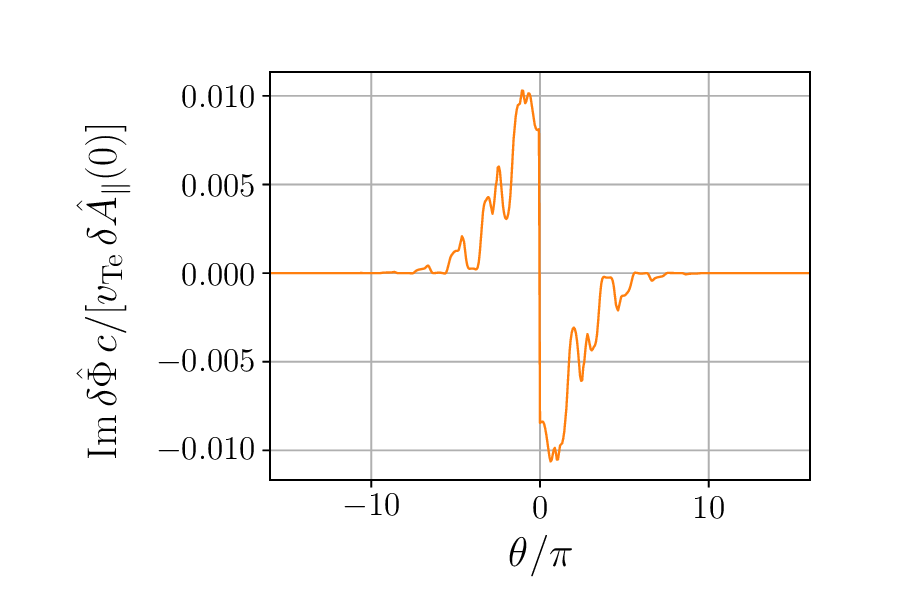}      \\
      \includegraphics[width=5.5cm]{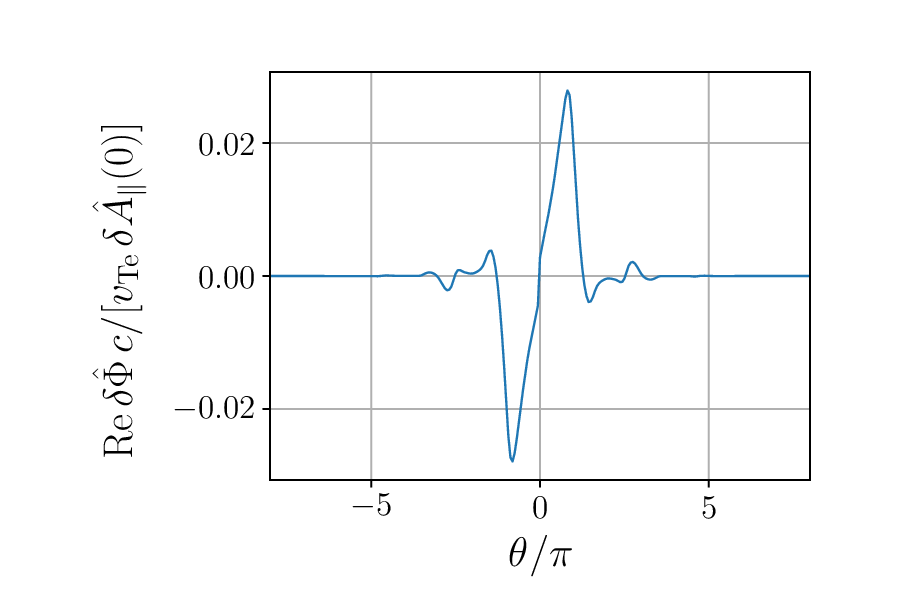}
      \includegraphics[width=5.5cm]{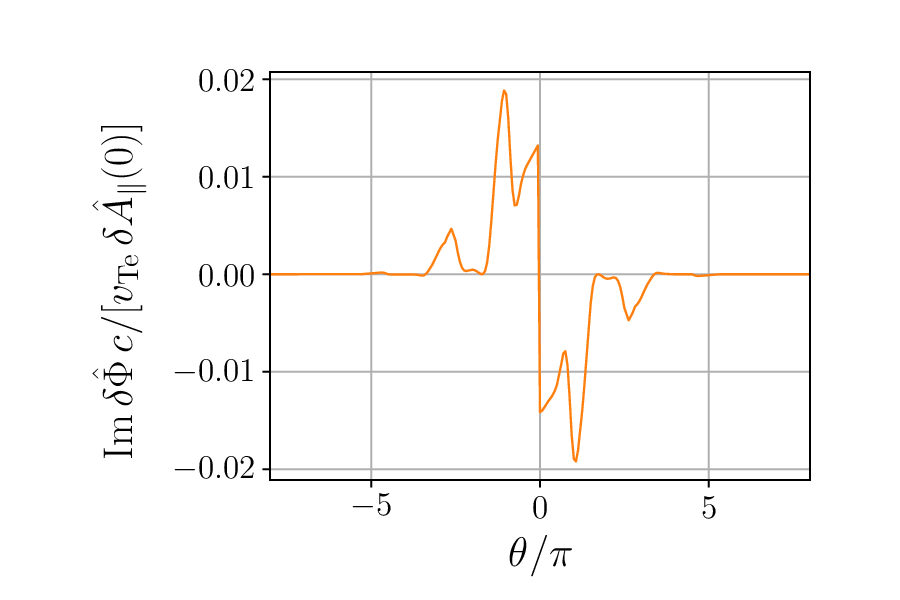}
  \caption{ The real (left) and imaginary (right) parts of the normalized fluctuating
    electrostatic potential $\delta \hat{\Phi}(\theta) c /[v_{\it T\rm e} \delta \hat{A}_{\parallel}(0)]$ when
$\theta_0=0$. From top to bottom, the five
    rows correspond to the toroidal mode numbers 25, 50, 100, 200,
    and 400, respectively, which can be converted to~$k_\wedge
    \rho_{\rm e}$ values by comparing the upper and lower horizontal
    axes in figure~\ref{fig:omega_k_wedge}.
    \label{fig:Phi} }
\end{center}
\end{figure}

In figure~\ref{fig:Phi}, we plot~$\delta \hat{\Phi}$ for $n=25$, 50, 100, 200, and~400, with $n$
increasing from the top row to the bottom row.  As $n$ increases, the
width $\Delta \theta$ of the~$\delta \hat{\Phi}$ eigenfunction
decreases, in agreement with the estimate in
Section~\ref{sec:collisionless}
that~$\Delta \theta \sim (k_\wedge \rho_{\rm e})^{-1}$.  Aside from
the decrease in~$\Delta \theta$, the qualitative shape of the
$\delta \hat{\Phi}(\theta)$ eigenfunction and the value
of~$\gamma /\omega_{\rm r}$ remain similar as~$n$ ranges from $25$
to~$400$, even though this range of~$n$ corresponds to
$k_\wedge \rho_{\rm i} $ values ranging from less than~1 to greater
than~1, where $\rho_{\rm i} $ is the ion gyroradius. This latter point is
consistent with the analysis of Section~\ref{sec:collisionless}, in
which the ions play no particular role in the MTM other than via their
Boltzmann response.

Although we have retained the trapped-electron $\hat{h}_{\rm e}$ terms
in our analysis, they have only a modest effect on the MTM growth rate
for the spherical-tokamak equilibrium that we investigated in
Section~\ref{sec:numerical}.  For example, these terms
increase~$\gamma$ by $\simeq 10 \%$ for the~$n=50$ data point
in~figure~\ref{fig:omega_k_wedge} and by~$\simeq 0.6\%$ for the
$n=400$ data point.

\section{Conclusion}
\label{sec:conclusion}

In this paper, we have derived the collisionless gyrokinetic MTM dispersion
relation, which is given by~(\ref{eq:disp0}),
\begin{equation}
   \omega - \omega_0 + i \pi^{1/2} \left(\frac{v_{\it T\rm e}}{L}+
     \frac{\omega^2 B_{\rm max}}{2 v_{\it T\rm e}} \int_{-\infty}^\infty \d \theta\,
     |J| \Gamma \, \delta \tilde{\Phi} \right) = 0,
\label{eq:disp0_conc} 
\end{equation} 
supplemented by the quasineutrality condition, (\ref{eq:QN}), which
determines the function~$\delta \tilde{\Phi}(\theta,\omega)$. In
agreement with past studies, we find that the MTM is driven unstable
by the electron temperature gradient rather than by the density
gradient (Section~\ref{sec:zero_eta_e}), and that the MTM instability
mechanism requires the electrostatic potential fluctuation to be present: when the
term containing~$\delta \tilde{\Phi}$ in~(\ref{eq:disp0_conc}) is
neglected, the imaginary part of~$\omega$ is strictly negative. The
instability mechanism also depends on magnetic drifts in a way that is
quantified by the~$\overline{ I}_0^\theta$ terms in (\ref{eq:disp0}),
(\ref{eq:defGamma}) and~(\ref{eq:QN}) through~(\ref{eq:defg}).

As discussed in Section~\ref{sec:Ampere}, (\ref{eq:disp0_conc}) is just the parallel component of
Ampere's law evaluated at~$\theta=0$.  Upon multiplication by
$C_2 = \sigma_J n_0 e^2 B v_{\it T\rm e}\hat{\psi}_{\parallel, \infty} /
(\pi^{1/2} T_{\rm e} B_{\rm max} \omega)$, (\ref{eq:disp0_conc})  becomes
\begin{equation} 
  \delta \hat{j}_{\delta E_\parallel} + \delta \hat{j}_{\delta B_\psi} -
  \delta \hat{j}_{\rm net} + \delta \hat{j}_{\delta \Phi}= 0,
  \label{eq:disp_j_conc}
\end{equation}
where $ \delta \hat{j}_{\delta E_\parallel}= C_2 \omega$ is the
parallel current density at~$\theta=0$ produced by the inductive
parallel electric field~$i c^{-1} \omega\, \delta \hat{A}_\parallel$,
$\delta \hat{j}_{\delta B_\psi} = - C_2 \omega_0$ is the parallel
current density at $\theta=0$ produced by
$ \delta \hat{B}_\psi = \delta \hat{\boldsymbol{B}} \cdot \nabla \psi
/ |\nabla \psi| = - i k_\wedge \delta \hat{A}_\parallel$ (i.e., by
electrons streaming along perturbed magnetic-field lines that wander
across the equilibrium flux surfaces), $\delta \hat{j}_{\rm net}$ is
the net parallel current density at $\theta=0$ (which is given by
$k_\perp^2 c\, \delta \hat{A}_\parallel(0) / 4\pi$, or, equivalently,
$-C_2$ times the term proportional to~$1/L$ in (\ref{eq:disp0_conc})),
and~$\delta \hat{j}_{\delta \Phi}$ is $C_2$ times the term
proportional to~$\delta \tilde{\Phi}$ in (\ref{eq:disp0_conc}), or,
equivalently, the parallel current density at~$\theta=0$ produced
by~$\delta \hat{\Phi}$, the nature of which is discussed at the end of
Section~\ref{sec:determination}.  Figure~\ref{fig:omega_k_wedge} shows
that~$\omega_{\rm r} \simeq \omega_0$ and
$\gamma/\omega_{\rm r} \lesssim 1/5$ for MTMs at
$k_\wedge \rho_{\rm e} \ll \beta_{\rm e}$ in the spherical-tokamak
equilibrium considered in Section~\ref{sec:numerical}, which implies
that the~$\delta \tilde{\Phi}$ term in (\ref{eq:disp0_conc}) is
significantly smaller than the~$\omega$ and~$\omega_0$ terms, and
hence that $\delta \hat{j}_{\delta \Phi}$ is significantly smaller
than the current densities~$\delta \hat{j}_{\delta E_\parallel}$
and~$\delta \hat{j}_{\delta B_\psi}$ that result
from~$\delta \hat{A}_\parallel$. The MTMs in
Section~\ref{sec:numerical} at
$k_\wedge \rho_{\rm e} \ll \beta_{\rm e}$ are thus basically magnetic
drift waves (with
$\omega \simeq \omega_0 \simeq \bm{k}_\perp \cdot \bm{v}_{\ast \rm e}$
--- see Section~\ref{sec:magneticdriftwaves} and the discussion of
figure~\ref{fig:omega_k_wedge} in Section~\ref{sec:numerical}) that
are driven weakly unstable ($\gamma \ll \omega_{\rm r}$) by the
electron temperature gradient via~$\delta \hat{\Phi}$.

The analogy to the magnetic drift wave suggests the following
heuristic way of thinking about how the MTM frequency is determined.
In the case of the magnetic drift wave
(Section~\ref{sec:magneticdriftwaves}), $\omega$ is fixed by requiring
that the force from the inductive parallel electric
field~$i c^{-1} \omega\, \delta \hat{A}_\parallel$ cancel out the
component of the electron pressure force parallel to the total
magnetic field~$\bm{B} + \delta \bm{B}$ in order to avoid unphysically
large currents.  Analogously, in the case of the MTM, $\omega$ is
determined by requiring that~$\delta \hat{j}_{\delta E_\parallel}$
offset any difference between~$\delta \hat{j}_{\rm net}$ and
$\delta \hat{j}_{\delta B_\psi} + \delta \hat{j}_{\delta \Phi}$.  In
an approximate sense (with corrections of
order~$\delta \hat{j}_{\delta \Phi}/\delta \hat{j}_{\delta B_\psi}$),
the real part of~$\omega$ is determined by the condition that the
$C_2 \omega_{\rm r}$ part of
$\delta \hat{j}_{\delta E_\parallel}$
cancel~$\delta \hat{j}_{\delta B_\psi}$, which is always~$90^\circ$
out of phase with~$\delta \hat{j}_{\rm net}$. This first condition is
very similar to the force balance that arises in the magnetic drift
wave and is the reason that the MTM propagates at approximately the
electron diamagnetic drift velocity~$\bm{v}_{\ast \rm e}$.  The
MTM growth rate is then determined by the condition that the
$C_2 \cdot i \gamma $ part of
$\delta\hat{j}_{\delta E_\parallel}$ match the difference
between~$\delta \hat{j}_{\rm net} $ and the part
of~$\delta \hat{j}_{\delta \Phi}$ that is in phase
with~$\delta \hat{j}_{\rm net}$.

There is some tension between the numerical examples presented in
Section~\ref{sec:numerical} and previous studies that suggest that
$\gamma \rightarrow 0$ 
as~$\nu_{\rm e} \rightarrow 0$ when $k_\wedge \rho_{\rm i} < 1$
\citep{applegate07,guttenfelder12a,patel22}, and further work is
needed to clarify the extent of, and reasons for, this
discrepancy. Other potentially fruitful directions for future research
include applying the results of this paper to
other tokamak equilibria and  stellarator equilibria
and generalizing the analysis 
to account for collisions.

\vspace{0.3cm} 
\noindent {\bf Acknowledgements}
\vspace{0.1cm}

We thank Ian Abel, Toby Adkins, Michael Barnes, Evan Chandran, Steve
Cowley, Bill Dorland, Michael Hardman, Per Helander, Daniel Kennedy,
Nuno Loureiro, Felix Parra, Bhavin Patel, and Yujia Zhang for helpful
discussions. BC thanks Merton College and the Rudolf Peierls Centre
for Theoretical Physics for their hospitality throughout his monthlong
visit to the University of Oxford in 2022, when this research project
was initiated.

\vspace{0.3cm} 
\noindent {\bf Funding}
\vspace{0.1cm} 

The work of AAS was supported in part by the UK EPSRC grant
EP/R034737/1 and by the Simons Foundation via a Simons Investigator
award.

\vspace{0.3cm} 
\noindent {\bf Declaration of Interests}
\vspace{0.1cm} 

The authors report no conflict of interest.

\vspace{0.3cm}
\noindent {\bf Author ORCID}
\vspace{0.1cm}

B. Chandran, https://orcid.org/0000-0003-4177-3328; A. Schekochihin, https://orcid.org/0000-0003-4421-1128

\appendix

\section{Trapped electrons}
\label{ap:trapped} 

To determine $\hat{h}_{\rm e\pm}$ for trapped electrons, we
integrate~(\ref{eq:gk}) for both $\hat{h}_{\rm e+}$ and
$\hat{h}_{\rm e -}$
from~$\theta$ 
to the nearest bounce points
surrounding~$\theta$, denoted $\theta_1$ and~$\theta_2$, with
$\theta_1 < \theta < \theta_2$. This yields four equations
for the six unknowns $\hat{h}_{\rm e+}(\theta)$, $\hat{h}_{\rm e-}(\theta)$,
$\hat{h}_{\rm e+}(\theta_1)$, $\hat{h}_{\rm e-}(\theta_1)$,
$\hat{h}_{\rm e+}(\theta_2)$, and $\hat{h}_{\rm e-}(\theta_2)$.
We eliminate two of these six unknowns, $\hat{h}_{\rm e+}(\theta_2)$, and $\hat{h}_{\rm e-}(\theta_2)$,
by imposing the trapped-particle boundary conditions
\begin{equation}
  \hat{h}_{\rm e +}(\theta_1) = \hat{h}_{\rm e -}(\theta_1) \qquad
    \hat{h}_{\rm e +}(\theta_2) = \hat{h}_{\rm e -}(\theta_2) ,
    \label{eq:trappedbc}
  \end{equation}
  leaving a system of four equations in four unknowns. Two 
  subtractions (of one equation from another) 
  eliminate $\hat{h}_{\rm e+}(\theta_1)$ and
  $\hat{h}_{\rm e-}(\theta_1)$, leaving a system of two linear equations in
  the two unknowns $\hat{h}_{\rm e+}(\theta)$ and
  $\hat{h}_{\rm e-}(\theta)$, whose solutions  can be combined to
  yield
\begin{equation}
  \begin{split}
    \frac{1}{2} \left[ \hat{h}_{\rm e + }(\theta) + \hat{h}_{\rm e -
      }(\theta)\right]_{\rm tr}
    =
    \frac{\xi_{\rm e}}{\sin I_{\theta_1}^{\theta_2}} \left[
        \int_{\theta_1}^\theta \d\theta^\prime
        \frac{JB}{|v_\parallel|}\left(
\sigma_1 \cos I_{\theta_1}^{\theta^\prime} \cos I_{\theta_2}^\theta +
i\sigma_2 \sin I_{\theta_1}^{\theta^\prime} \cos I_{\theta_2}^\theta
\right)\right. \\
\left.+ \int_\theta^{\theta_2} \d \theta^\prime \frac{JB}{|v_\parallel|}
\left( \sigma_1 \cos I_{\theta_2}^{\theta^\prime} \cos
  I_{\theta_1}^\theta + i\sigma_2 \sin I_{\theta_2}^{\theta^\prime}
  \cos I_{\theta_1}^\theta \right)\right]
  \end{split}
\label{eq:trappedeven} 
\end{equation}
and  
\begin{equation}
  \begin{split}
    \frac{1}{2} \left[ \hat{h}_{\rm e + }(\theta) - \hat{h}_{\rm e -
      }(\theta)\right]_{\rm tr}
    =
    \frac{\xi_{\rm e}}{\sin I_{\theta_1}^{\theta_2}} \left[
        \int_{\theta_1}^\theta \d\theta^\prime
        \frac{JB}{|v_\parallel|}\left(
i\sigma_1 \cos I_{\theta_1}^{\theta^\prime} \sin I_{\theta_2}^\theta -
\sigma_2 \sin I_{\theta_1}^{\theta^\prime} \sin I_{\theta_2}^\theta
\right)\right. \\
\left.+ \int_\theta^{\theta_2} \d \theta^\prime \frac{JB}{|v_\parallel|}
\left( i \sigma_1 \cos I_{\theta_2}^{\theta^\prime} \sin
  I_{\theta_1}^\theta - \sigma_2 \sin I_{\theta_2}^{\theta^\prime}
  \sin I_{\theta_1}^\theta \right)\right],
  \end{split}
\label{eq:trappedodd} 
\end{equation}
where   $\sigma_1 \equiv J_0(\alpha_{\rm e}) \,\delta \hat{\Phi} $ and
$ \sigma_2 \equiv |v_\parallel|   J_0(\alpha_{\rm e})
  \,\delta \hat{A}_\parallel/c $.\footnote{If we had retained $\delta \hat{B}_\parallel$ in
  (\ref{eq:gk}), then we would have had to add $(v_\perp/ k_\perp c)J_1(\alpha_{\rm e})\, \delta
  \hat{B}_\parallel$ to~$\sigma_1$.}
Equations~(\ref{eq:trappedeven}) and
(\ref{eq:trappedodd}) were obtained by \cite{tang80}, but with
two minor differences; the
$i\sigma_2 \sin I_{\theta_2}^{\theta^\prime} \cos I_{\theta_1}^\theta$ term
at the end of (\ref{eq:trappedeven}) was written as
$i\sigma_2 \sin I_{\theta_2}^{\theta^\prime} \sin I_{\theta_1}^\theta$,
 and the
$i \sigma_1 \cos I_{\theta_1}^{\theta^\prime} \sin
I_{\theta_2}^\theta$ term at the beginning of (\ref{eq:trappedodd})
was written as
$i \sigma_1 \cos I_{\theta_1}^{\theta^\prime} \sin
I_{\theta_1}^\theta$.

When we evaluate the trapped-electron contribution to the parallel
component of Ampere's law in Section~\ref{sec:Ampere}, we need only
the odd (in $v_\parallel$) part of~$\hat{h}_{\rm e}$,
given by~(\ref{eq:trappedodd}), and we only consider
$\theta \sim {\rm O}(1)$, where all quantities of the form
$I_{\theta_a}^{\theta_b}$ in (\ref{eq:trappedodd})
are~$\sim {\rm O}(k_\wedge \rho_{\rm e} ) \ll 1$.  The smallness of
the~$I_{\theta_a}^{\theta_b}$ terms makes the trapped-electron
contribution to~$\hat{h}_{\rm e+ } - \hat{h}_{\rm e-}$ a
factor~$\sim k_\wedge \rho_{\rm e}$ smaller than the passing-electron
contribution.\footnote{This point is further clarified by the
 estimate given in Appendix~\ref{ap:justification}  
 of terms $\circled{2a}$ and~$\circled{2b}$ in (\ref{eq:ampere0}).}
 Physically, magnetic mirroring reduces the part of the
distribution function that is odd in~$v_\parallel$, making the
trapped-electron parallel current negligible.

In Section~\ref{sec:quasineutrality}, when we evaluate the
quasineutrality condition, we need only the even part
of~$\hat{h}_{\rm e}$, given by~(\ref{eq:trappedeven}), and we consider
only $|\theta|  \gg 1$, as the value
of~$\delta \hat{\Phi}$ at $|\theta| \ll (k_\wedge \rho_{\rm e})^{-1}$
contributes only a small correction to~(\ref{eq:disp0}). At 
$|\theta| \gg 1$, $\delta \hat{A}_\parallel$ is negligible because
of~(\ref{eq:c1}), and we can drop the $\sigma_2$ term
in~(\ref{eq:trappedeven}).  If $\theta_a$ or $\theta_b$ differs from a
bounce point in any of the $ I_{\theta_a}^{\theta_b}$ terms in
(\ref{eq:trappedeven}), then the dominant contribution to that
$I_{\theta_a}^{\theta_b}$ term comes from the $\omega_{\rm De}$ term
in~(\ref{eq:defI}), and in particular the part of~$\omega_{\rm De}$
that arises from component of~$\bm{v}_{\rm De}$ along~$\nabla \psi$,
which satisfies the identity~\citep{hinton85}
\begin{equation}
  \boldsymbol{v}_{\rm De} \cdot \nabla \psi = \frac{v_\parallel I(\psi)}{JB}
  \left.\frac{\partial }{\partial \theta}\right|_{\alpha, \psi, E, \mu}
  \left(\frac{v_\parallel}{\Omega_{\rm e}}\right).
  \label{eq:identity}
\end{equation}
At $|\theta| \gg 1$, (\ref{eq:kpsi})  implies that
$k_{\perp \psi} = - n |\nabla \psi| ({\rm d} q / {\rm d}\psi) \theta$
to leading order in~$1/\theta$ (where we have taken~$\theta_0 \sim {\rm
  O}(1)$). We thus find that 
\begin{equation}
I_{\theta_a}^{\theta_b} = n q^\prime(\psi) I(\psi) \left. \frac{\theta
  |v_\parallel|}{\Omega_{\rm e}}\right|_{\theta_a}^{\theta_b}
\label{eq:Iapprox} 
\end{equation} 
to leading order in~$1/\theta$,  where
$\theta_a$ and $\theta_b$ are chosen from $\{\theta_1, \theta_2,
\theta, \theta^\prime\}$, provided all three of the following
conditions are met: (1) $|\theta| \gg 1$;
(2) $\theta_a$ or~$\theta_b$ is not a bounce point; and (3) $|\theta_a - \theta_b| \sim {\rm
  O}(1)$. When $\theta_a$ and $\theta_b$ are both bounce points, the
right-hand side of~(\ref{eq:Iapprox}), which is usually the dominant
term in $I_{\theta_a}^{\theta_b}$, vanishes, and
$I_{\theta_a}^{\theta_b}$ is instead dominated by the remaining terms, which,
although non-vanishing, are~$\ll 1$. We can thus write
\begin{equation}
\sin I_{\theta_1}^{\theta_2} = I_{\theta_1}^{\theta_2} + \dots =
\int_{\theta_1}^{\theta_2}{\rm d}\theta\, \frac{JB}{|v_\parallel|} \left(\omega -
  \omega_{\rm De}\right) + \dots = \sigma_J \tau_{\rm b } \left\langle \omega
  - \omega_{\rm De} \right\rangle_{\rm b}  + \dots,
\label{eq:Ibounce} 
\end{equation} 
where $\dots$ represents higher-order corrections, and
\begin{equation}
\langle \dots \rangle_{\rm b} \equiv \frac{1}{\tau_{\rm
    b}}\int_{\theta_1}^{\theta_2} {\rm d}\theta\,\frac{|J|B}{|v_\parallel|} (\dots)
\qquad \mbox{and} \qquad
\tau_{\rm b}(E, \mu) \equiv \int_{\theta_1}^{\theta_2}{\rm d}\theta\,
\frac{|J|B}{|v_\parallel|}
\label{eq:tau_b} 
\end{equation}
are the bounce average and bounce time, respectively. 
Upon substituting (\ref{eq:Iapprox}) and~(\ref{eq:Ibounce})
into~(\ref{eq:trappedeven}), we find that 
\begin{equation}
      \left[ \hat{h}_{\rm e + }(\theta) + \hat{h}_{\rm e -
        }(\theta)\right]_{\rm tr}
    = \frac{ 2\xi_{\rm e}\cos\left(P\right)}{\left\langle \omega - \omega_{\rm
          De}\right\rangle_{\rm b}} 
\left\langle J_0(\alpha_{\rm e}) \cos\left(P\right) \delta \hat{\Phi}
\right\rangle_{\rm b},
\label{eq:trappedeven2}
\end{equation} 
where $P = n q^\prime(\psi) I(\psi) |v_\parallel|
    \theta/\Omega_{\rm e}$. Substituting (\ref{eq:trappedeven2})
    into~(\ref{eq:quasineutrality}) for the value of $\hat{h}_{\rm e+} +
    \hat{h}_{\rm e-}$ at
    $\mu > E/B_{\rm max}$ leads to the $W_{\rm tr}(\theta, \theta^\prime)$
    term in~(\ref{eq:QN}).

\section{Justification of (\ref{eq:comp}) }
\label{ap:justification}

By the same arguments that led to (\ref{eq:disp0}), the sum of terms
$\circled{3a}$
and $\circled{3b}$ in (\ref{eq:ampere0}) is
\begin{equation}
  \circled{3a} + \circled{3b} =  
 \frac{ \sigma_J n_0 e^2v_{\it T\rm e}(\omega- \omega_0) \hat{\psi}_{\parallel, \infty}}{ \pi^{1/2} T_{\rm
      e}B_{\rm max}\omega}
  \label{eq:3aplus3b}.
  \end{equation} 
Equations~(\ref{eq:c1})  and~(\ref{eq:beta_e_min}) imply that
\begin{equation}
  \circled{1} = C_1 = -\,\frac{ 2i \sigma_J  n_0 e^2 \hat{\psi}_{\parallel, \infty}}{m_{\rm e} \omega
    B_{\rm max} L}
  \label{eq:c1_psi_par}.
\end{equation}
The second line of (\ref{eq:comp}) follows from dividing
(\ref{eq:c1_psi_par}) by~(\ref{eq:3aplus3b}) and making use of the
orderings in~Section~\ref{sec:ordering}.

To estimate terms $\circled{2a}$ and~$\circled{2b}$ in
(\ref{eq:ampere0}), we first estimate~$I_0^\theta$ in~(\ref{eq:defI})
for electrons with $v \sim v_{\it T\rm e}$.
Equations (\ref{eq:omegaordering}), (\ref{eq:n_ordering}),
and~(\ref{eq:omegaDs})  imply that
the contributions to~$I_0^\theta$ from terms proportional to $\omega $
or $ k_\wedge$ are~$\sim k_\wedge \rho_{\rm e}
\theta$. The remaining contribution to~$I_0^\theta$, which is
$\propto k_{\perp \psi}$, is a little trickier to estimate,
because~$k_{\perp \psi}$ grows secularly with~$\theta$.  Keeping just
the dominant part of~$k_{\perp \psi}$ at large~$|\theta|$, which
from~(\ref{eq:kpsi}) is
$- n |\nabla \psi| ({\rm d} q / {\rm d}\psi) \theta$,
employing~(\ref{eq:identity}),
and defining $v_{{\rm De},\psi} = \bm{v}_{D\rm e} \cdot \nabla
\psi/|\nabla \psi|$, we find that
\begin{equation}
   \int_0^\theta \, {\rm d}\theta^\prime \,\frac{JB}{|v_\parallel|}
   n |\nabla \psi| \frac{{\rm d} q}{{\rm d}\psi} \theta^\prime v_{{\rm De}, \psi}  = n  \frac{{\rm d} q}{{\rm d}\psi}  I(\psi) \theta \left(
   \frac{|v_\parallel|}{\Omega_{\rm e}} - \left \langle
     \frac{|v_\parallel|}{\Omega_{\rm e}}\right\rangle \right) \sim
k_\wedge \rho_{\rm e} \theta,
 \label{eq:omegaDepsi}
 \end{equation} 
where $\langle f \rangle  \equiv (1/\theta)\int_0^\theta {\rm
  d}\theta^\prime \, f(\theta^\prime)$. Therefore, $I_0^\theta$ in its
entirety is~$\sim k_\wedge \rho_{\rm e}\theta$.

At $|\theta| \lesssim (k_\wedge \rho_{\rm e})^{-1}$,
$|I_0^\theta| \lesssim 1$,
$e^{\pm i \overline{ I}_0^\theta} \sim {\rm O}(1)$, and
$J_0(\alpha_{\rm e}(\theta)) \sim {\rm O}(1)$.
Equations~(\ref{eq:defGamma}) and~(\ref{eq:defW}) then show,
respectively, that $\Gamma(\theta) \sim {\rm O}(1)$ and
$W_{\rm p}\left(\theta, \theta^\prime\right) \sim k_\wedge \rho_{\rm e}$ when
$|\theta| \lesssim (k_\wedge \rho_{\rm e})^{-1}$
and $|\theta^\prime| \lesssim (k_\wedge \rho_{\rm e})^{-1}$. On the other hand, at
$|\theta| \gg (k_\wedge \rho_{\rm e})^{-1}$, $\overline{ I}_0^\theta$
becomes large, and $\exp\left(\pm i \overline{ I}_0^\theta\right)$
becomes a rapidly oscillating function of velocity.  This rapid
oscillation and the decay in $J_0(\alpha_{\rm e}(\theta))$
at~$|\theta| \gg (k_\wedge \rho_{\rm e})^{-1}$
cause~$|\Gamma(\theta)|$ to
decay to values~$\ll 1$ 
and~$|W_{\rm p}\left(\theta, \theta^\prime\right)|$ to
decay to values~$\ll k_\wedge \rho_{\rm e}$ when~$|\theta|$ or
$|\theta^\prime|$ is~$\gg (k_\wedge \rho_{\rm e})^{-1}$.
For similar reasons, $W_{\rm tr}\left(\theta ,\theta^\prime\right) \sim {\rm O}(1)$ when
$|\theta| \lesssim (k_\wedge \rho_{\rm e})^{-1}$, and
$W_{\rm tr}(\theta ,\theta^\prime) \ll 1$ when
$|\theta|$ or~$|\theta^\prime|$ is~$\gg (k_\wedge \rho_{\rm e})^{-1}$.
Given these approximate values of~$W_{\rm p}\left(\theta,
  \theta^\prime\right)$, $\Gamma$, and
$W_{\rm tr}(\theta, \theta^\prime)$, it follows from~(\ref{eq:QN}) that $\delta \tilde{\Phi} \sim {\rm O}(1)$ at
$|\theta | \lesssim (k_\wedge \rho_{\rm e})^{-1}$ and that
$|\delta \tilde {\Phi}| \ll 1$ at $|\theta| \gg (k_\wedge \rho_{\rm e})^{-1}$.  This behaviour of
$\delta \tilde{\Phi}$ enables us to estimate the magnitude of terms
$\circled{2a}$ and $\circled{2b}$ in (\ref{eq:ampere0}) by setting
$\delta \hat{\Phi}(\theta^\prime) \sim \hat{\psi}_{\parallel, \infty}$
for $|\theta^\prime| \lesssim (k_\wedge \rho_{\rm e})^{-1}$ and
$\delta \hat{\Phi}(\theta^\prime) \rightarrow 0 $ for
$|\theta^\prime| \gg (k_\wedge \rho_{\rm e})^{-1}$, which then, in
conjunction with (\ref{eq:3aplus3b}), leads to the first line
of~(\ref{eq:comp}).

\bibliography{articles}

\bibliographystyle{jpp}

\end{document}